\documentclass[VANCOUVER,STIX1COL]{WileyNJD-v2}

\newcommand{\e}[1]{\mbox{\lstinline[basicstyle=\normalsize]|#1|}}
\usepackage{boogie}
\usepackage{eiffel}
\usepackage{listings}
\usepackage{graphicx}
\usepackage{url}
\usepackage{makecell}
\usepackage{paralist}
\usepackage{subfigure}
\usepackage{comment}
\usepackage{xcolor}
\renewenvironment{comment}{}{}
\usepackage{bbding}
\usepackage[]{graphicx}
\usepackage{scalerel}
\usepackage{algpseudocode}
\usepackage{algorithm}
\usepackage{amsmath}
\usepackage{colortbl}
\usepackage[]{hyperref}

\makeatletter
\@addtoreset{chapter}{part}
\makeatother

\newcommand{\tabincell}[2]{\begin{tabular}{@{}#1@{}}#2\end{tabular}}

\usepackage{amsfonts, amsmath, amssymb}

\newcommand{\gt}[1]{{\texttt{#1}}}

\usepackage{fontawesome}

\articletype{Research article}%

\received{26 April 2016}
\revised{6 June 2016}
\accepted{6 June 2016}

\begin{document}

\title{Exploiting the Complementarity of Proofs and Tests}
\title{Generating a failing test from a failed proof}
\title{Using a failed proof to generate a failing test}
\title{A failed proof can yield a useful test}

\author{Li Huang}

\author{Bertrand Meyer}

\authormark{LI HUANG, BERTRAND MEYER}

\address{\orgdiv{Chair of Software Engineering}, \orgname{Constructor Institute}, \orgaddress{\state{Schaffhausen}, \country{Switzerland}}}


\corres{Li Huang, Chair of Software Engineering, Constructor Institute, Rheinweg 9, 8200 Schaffhausen, Switzerland.
\email{li.huang@sit.org}}



\abstract{A successful automated program proof is, in software verification, the ultimate triumph. In practice, however, the road to such success is paved with many failed proof attempts. Unlike a failed test, which provides concrete evidence of an actual bug in the program, a failed proof leaves the programmer in the dark. Can we instead learn something useful from it?

The work reported here takes advantage of the rich information that some automatic provers internally collect about the program when attempting a proof. If the proof fails, the Proof2Test tool presented in this article uses the counterexample generated by the prover (specifically, the SMT solver underlying the Boogie tool used in the AutoProof system to perform correctness proofs of contract-equipped Eiffel programs) to produce a failed test, which provides the programmer with immediately exploitable information to correct the program.
The discussion presents Proof2Test  and the application of the ideas and tool to a collection of representative examples.}
\begin{comment}
\end{comment}


\maketitle

\section{Introduction}
\label{intro}

We usually seek success, but failures can be useful too. In software verification, for example, a failed \textit{test} provides the programmer with concrete, immediately actionable information about a bug, helping to fix it.

A failed attempt at a \textit{proof}, on the other hand, seems at first not to help much; it does not even demonstrate that there is a bug (and even less what that bug might be), as it could simply reflect a problem with the proof process.

The work reported here takes advantage of the power of a modern prover to make a failed proof useful too, by  deriving a failed test from it. Examples handled so far seem to confirm the potential of the approach.

\subsection{Tests and proofs}

This work combines tests and proofs, a combination whose benefits was not  always self-evident. In fact,
tests and proofs have long been considered the warring siblings of software verification. They are characterized by dual benefits and limitations:

\begin{itemize}
    \item
    Tests have, in their favor, their concreteness and relative ease of preparation. They face, however, a fundamental limitation: successful tests, regardless of how many of them, do not guarantee the correctness of software.
    \item
    Proofs do hold that promise of correctness, \textit{when they succeed}. In practice, however, the path towards a proof is arduous, involving many intermediate steps at which the proof fails and it is hard to know what needs to be done, such as changing the program, changing the specification, or finding a way to overcome a limitation of the proof tool.
\end{itemize}

\begin{table*}[htbp]
  \centering
   \renewcommand\arraystretch{1.3}
    \begin{tabular}{|c|c|c|}
    \hline
      & Proof & Test \\
    \hline
    Success &  \tabincell{c}{  \faThumbsOUp\  Program is correct} &  \tabincell{c}{  {\faThumbsODown} \ Program works for \textit{one} more case}
    \\ \hline
    Failure &  \cellcolor{green!25}{\tabincell{c}{{\faThumbsODown}\  Inconclusive, don't know what's wrong}} &  \tabincell{c}{\faThumbsOUp \  Program is incorrect}
    \\ \hline
    \end{tabular}%
  \caption{The different meanings of failure and success for proofs and for tests}
\label{proof_and_test}
\end{table*}%

 \noindent Can the two techniques help each other by combining their respective advantages? This article proposes a specific answer, exploiting the observation that  \textit{failure} has different consequences for proofs and for tests. Table \ref{proof_and_test} shows the difference. \faThumbsOUp\ denotes the interesting cases, those in which we can conclusively deduce a useful property:
 \begin{itemize}
     \item
 Top-left entry: proof succeeds (great news, the program is correct).
 \item
 Bottom-right entry: test fails (program is incorrect, we have found a bug, less triumphal news but still concrete and useful).
  \end{itemize}
\noindent The other two cases, marked \faThumbsODown, are disappointing since they offer little practical value:
\begin{itemize}
    \item
A successful test (top right) does not tell us that the program is correct in the general case. It is just one drop of correctness in the ocean of possible cases.
\item
A failed proof (bottom left) does not tell us whether the problem is with the program, the specification, or the limitations of the proof tool itself.
\end{itemize}

\noindent This last case, a failed proof, shaded in the table, is particularly frustrating. While a mechanically-supported program proof is the most exciting prospect in the search for program correctness, the daily practice of attempting to prove programs consists of getting notices of proof failure and attempting to remove the cause for the failure. It is not so different psychologically from the practice of \textit{debugging} a program, except that it is a static rather than dynamic form of debugging, using only the program text and no executions.

Being static, this process lacks the immediacy and concreteness of a test, which (when it fails, as in the bottom-right entry) gives the programmer a direct example of an input that causes the program to violate its specification. Programmers can exploit such a test case as an immediately understandable indication of what needs to be fixed. In contrast, a failed proof just fails, giving you little insight as to what is actually wrong --- without even guaranteeing that the fault lies with the program (as noted, it could be in the specification, or we might just have hit a limitation of the proof technology). The downside is that the tester (a term meant here to denote a person, not a tool) has to devise the failing test case, often a difficult endeavor for a program that has already achieved a first level of approximate correctness.

Proof tools, on the other hand, accumulate considerable knowledge about the program, including when they fail to prove it. The work reported in this article consists of using that knowledge to go automatically from the disappointing bottom-left case to the directly useful bottom-right case, a failed test, with its exploitable example of program incorrectness.
\subsection{Proof2Test}

The tool presented in this article, Proof2Test, automatically generates failing tests from failed proofs. The underlying prover is AutoProof\cite{tschannen2015autoproof, autoproof}, the verifier for checking functional correctness of Eiffel\cite{meyer1997object} programs using axiomatic (Hoare-style) semantics.
The functional correctness of Eiffel programs is specified by \emph{contracts} (such as pre- and postconditions, class and loop invariants). When a proof fails, AutoProof's back-end prover (the Z3 solver\cite{de2008z3}, used through Boogie \cite{barnett2005boogie, le2011boogie}) will produce a counterexample. Proof2Test exploits the information in the counterexample to generate a concrete test case. The generated test case is directly exploitable and executable in AutoTest\cite{autotest, wei2010satisfying}, a testing tool for Eiffel programs based on run-time assertion checking.

Since AutoProof and AutoTest use the same contracts for program analysis and they are both integrated in the same development environment (Eiffel IDE), executing the test produced by Proof2Test from the failed proof can reproduce an analogous test failure in AutoTest: the execution will raise the run-time violation of the same failed contract. The general aim is to combine the benefits of both static and dynamic forms of verification:  proofs (performed by AutoProof) and tests (performed by AutoTest).

The source code for Proof2Test and examples, as well as complementary material, are available in a GitHub repository\footnote{\url{https://github.com/huangl223/Proof2Test}}. The repository contains a detailed readme.md file explaining how to run the examples and reproduce our results (or try new examples).


\subsection{Structure of the discussion}

To give the reader a general understanding of the approach before introducing the theory,     Section \ref{example} illustrates an example session of using Proof2Test. Section \ref{technology_stack} introduces the fundamental technologies used in the Proof2Test verification framework. Section \ref{implementation} describes the details of the implementation of Proof2Test. Section \ref{diagnosis} demonstrates how Proof2Test helps diagnose failed proofs.
Section \ref{experiment} evaluates the applicability of Proof2Test through a series of examples. After a review of related work in Section \ref{related_work}, Section \ref{limitations} describes limitations of the current state of the work and Section \ref{conclusion} concludes an analysis of limitations and ongoing work to overcome them.

For readers interested in detailed results, the \hyperref[appendix]{Appendix} covers the full data for two of the sample classes; it is extracted from a technical report \cite{testdetails}, complementing the present article by covering all samples.
\section{An example session with Proof2Test}
\label{example}

Before exploring the principles and technology behind the Proof2Test tool, we look at the practical use of the tool on a representative example (Figure \ref{listing:max}).

\begin{figure}
    \centering
\begin{lstlisting}[captionpos=b, basicstyle=\fontsize{0.3cm}{0.3cm}]
class MAX_IN_ARRAY
feature
    max (a: SIMPLE_ARRAY [INTEGER]): INTEGER	
	        require
			          a.count > 0
		      local	
		            i: INTEGER
		      do
			          from
			              Result := a [1]; i := 2
			          invariant
			              2 <= i and i <= a.count + 1
			              $\forall$ j: $1$ $|..|$ (i - 1) | a [j] $<=$ Result
			              $\exists$ j: $1$ $|..|$ (i - 1) | a [j] $=$ Result
			          until
				            i = a.count
			          loop
				            if a [i] > Result then
					                Result := a [i]
				            end
				            i := i + 1
                variant
                    a.count - i
			          end
		      ensure
			          result_is_maximum: $\forall$ j: $1$ $|..|$ a.count | a [j] $<=$ Result
			          result_in_array: $\exists$ j: $1$ $|..|$ a.count | a [j] $=$ Result
		      end
end
\end{lstlisting}
\caption{\e{MAX_IN_ARRAY} is a class with a function \e{max} which finds the maximum element of an integer array \e{a}; the type of \gt{a} is \e{SIMPLE\_ARRAY}, a library class providing simple array features}
\label{listing:max}
\end{figure}

The intent of the \e{max} function in class \e{MAX_IN_ARRAY} is to return into \e{Result} the maximum element of an integer array \e{a} of size \e{a.count}. The two postcondition clauses in lines 26 and  27 (labeled \emph{result\_is\_maximum} and \emph{result\_in\_array}) specify this intent:  every element of the array should be less than or equal to \e{Result}; and at least one element should be equal to \e{Result}.

When we try to verify the \e{max} function using AutoProof, verification fails and AutoProof returns the error message ``Postcondition \emph{result\_is\_maximum} (line 26) might be violated''. Such a generic message tells us that the prover cannot establish the postcondition, but does not enable us to find out why. We are left with hypotheses, including: incorrect loop initialization (line 10); incorrect loop exit condition (line 16); incorrect instruction in the loop body (lines 18 -- 21); incorrect loop invariant (lines 12 -- 14); or limitations of the prover itself (in other words, it might be that the program is correct and AutoProof is unable to prove it).

What a programmer typically would like to see in such a case is not a general negative result, stating that the prover cannot establish a property,  but a concrete, specific result, showing that a certain input breaks the specification. In other words, a failing test. Producing such a test is the purpose of Proof2Test. Figure \ref{listing:test_max} shows the generated test, which includes the following sequence of steps:
\begin{itemize}[\textbullet]
\item Create an instance \e{current\_object} of class \e{MAX_IN_ARRAY} (line 7).
\item Create an integer array \e{a} and fill it with values \e{0} at position \e{1} and \e{8} at index \e{2} (lines 8 -- 9).
\item Call the erroneous function \e{max} on \e{current\_object} with \e{a} as argument (line 10).
\end{itemize}

\begin{figure}
    \centering
    \begin{lstlisting}[basicstyle=\fontsize{0.3cm}{0.3cm}]
test_MAX_IN_ARRAY_max_1
	  local
		    current_object: MAX_IN_ARRAY
		    a: SIMPLE_ARRAY [INTEGER]
		    max_result: INTEGER
	  do
		    create current_object.default_create
		    create a.make (0)
		    a.force (0, 1); a.force (8, 2)
		    max_result := current_object.max (a)
	  end
\end{lstlisting}
    \caption{A test case of \e{max} with input argument \e{a[1] = 0}, \e{a[2] = 8}}
    \label{listing:test_max}
\end{figure}

\noindent Running this Proof2Test-generated test in AutoTest (the test framework for Eiffel) leads to a run-time failure where the problematic postcondition \emph{result\_is\_maximum} is violated. To explore the reason, we may step through the call \e{current\_object.max (a)}:
\begin{enumerate}
\item The precondition (line 5 in Figure \ref{listing:max}) evaluates to \e{True} with \e{a.count} = 2.

\item After the loop initialization (line 10), \e{Result} = 0 and \e{i} = 2.

\item All the loop invariants (lines 12 -- 14) are evaluated as \e{True} with \e{i} = 2, \e{a.count} = 2, \e{a[1]} = 0, \e{a[2] = 8}, \e{Result} = 0.

\item The exit condition of the loop (line 16) evaluates to \e{True} with \e{a.count} = 2 and \e{i} = 2, forcing the loop to terminate.

\item Postcondition \emph{result\_is\_maximum} (line 26) evaluates to \e{False} with \e{a [1]} = 0, \e{a [2]} = 8 and \e{Result} = 0, which causes AutoTest to report a run-time exception of postcondition violation.
\end{enumerate}

\noindent This execution of the generated test reveals the problem: on the array created by the test, which holds its maximum value (8) in the last element (at index 2 in \e{a}), the loop terminates too early, preventing the program from getting to that value.

 This scenario and its result provide the programmer with immediate evidence (not available from the failed proof attempt) of what is wrong with the program version of Figure \ref{listing:max}. Any competent programmer will see right away what needs to be done to eliminate the error: update the exit condition \e{i} \e{$=$}\ \e{a.count} (line 16) to permit one more loop iteration, changing it to \e{i}  \e{$>$} \ \e{a.count}.





This example illustrates how tests and proofs can be complementary techniques: while a successful proof conclusively shows that the program satisfies the given specification, a failed proof does not by itself tell us what is wrong with the program; in that case a test can bring the concrete evidence making it possible to proceed with the development process. The remaining sections explain the technology behind this approach.

\section{Technology stack}
\label{technology_stack}
Fundamental technologies used in Proof2Test include the Eiffel programming language, its libraries, the EiffelStudio IDE (development environment including compiler and AutoTest test framework) and two analysis tools, AutoProof (static) and AutoTest (dynamic).

\vspace{0.05in}
\subsection{Eiffel}
\label{eiffel}

The Eiffel object-oriented design and programming language\cite{meyer1997object, bertrand2016touch}  natively supports the Design by Contract\cite{meyer1992applying} methodology. An Eiffel program consists of a set of classes. A class represents a set of run-time objects characterized by the features available on them.
Figure \ref{listing:ACCOUNT} shows an Eiffel class representing bank accounts, devised for demonstration purposes and intentionally seeded with errors.

A class contains two kinds of features:
\emph{attributes} that represent data items associated with instances of the class, such as \e{balance} and \e{credit\_limit};
\emph{routines} representing operations applicable to these instances, including \e{make}, \e{available\_amount}, \e{deposit}, \e{withdraw}, \e{transfer}.
Routines are further divided into procedures (routines that have no returned value) and functions (routines that return a value). Here, \e{available\_amount} is a function returning an integer (denoted by a special variable \e{Result}), and the other routines are procedures.

A class becomes a \emph{client} of \e{ACCOUNT} by declaring one or more entities (attributes, local variables, formal arguments...) of type \e{ACCOUNT}, as in \e{a: ACCOUNT}.

The value of an reference entity is void initially and can be attached to an object by using the \e{create} instruction or through assignment of a non-void value. For example, \e{create a.make}($-$100) creates a new object whose \e{credit\_limit} is $-$100 and attaches it to the entity \e{a}.

\begin{figure}
    \centering
\noindent\begin{minipage}{.5\textwidth}
\begin{lstlisting}[basicstyle=\fontsize{0.3cm}{0.3cm}]
class ACCOUNT create
    make
feature
    make (limit: INTEGER)
            -- Initialize with credit limit `limit'.
        require
            limit <= 0
        do
            balance := 0 ; credit_limit := limit
        ensure
            balance_set: balance = 0
            credit_limit_set: credit_limit = limit
		end
	
    balance: INTEGER
        -- Balance of this account.
	
    credit_limit: INTEGER
        -- Credit limit of this account.

    available_amount: INTEGER	
        -- Amount available on this account.
        do
            Result := balance - credit_limit
        end

    deposit (amount: INTEGER)	
        -- Deposit `amount' into this account.
        do
            balance := balance + amount
        ensure
            balance_increased: balance > old balance
        end
\end{lstlisting}
\end{minipage}
\begin{minipage}{.45\textwidth}
\begin{lstlisting}[firstnumber=last, basicstyle=\fontsize{0.3cm}{0.3cm}]
withdraw (amount: INTEGER)	
    -- Withdraw `amount' from this bank account.
    require
        amount >= 0
        amount <= available_amount
    do
        balance := balance + amount
    ensure
        balance_set: balance = old balance - amount
    end

transfer (amount: INTEGER; other: ACCOUNT)	
    -- Transfer `amount' to `other'.
    require
        amount >= 0
        amount <= available_amount
    do
        balance := balance - amount
        other.deposit (amount)
    ensure
        withdrawal_made: balance = old balance - amount
        deposit_made: other.balance = old other.balance + amount
    end

invariant 		
    available_amount >= 0
end
\end{lstlisting}
\end{minipage}
\caption{A class (with bugs) intended to implement the behavior of bank accounts}
    \label{listing:ACCOUNT}
\end{figure}

Programmers can specify the properties of classes and features by equipping them with contracts of the following kinds:
\begin{itemize} [\textbullet]
\item Precondition (\e{require}): property that \emph{clients} must satisfy whenever they call a routine;  the precondition of \e{make} (line 6), for example, requires the value of \e{limit} to be negative or 0.
\item Postcondition (\e{ensure}): property that the routine (the \emph{supplier}) guarantees on routine exit, assuming the precondition and termination; for example, the postcondition of \e{deposit} states that the value of \e{balance} at the exit of the routine will be greater than its entry value (\e{old balance}).
\item Class invariant (\e{invariant}): constraint that must be satisfied by instances of the class after object creation and after every qualified call \e{x.r (args)} to a routine of the class; for example, the class invariant of \e{ACCOUNT} (line 58)  constrains the value of \e{available\_amount} to be always non-negative.
\item Loop invariant (\e{invariant}): property that the loop guarantees after initialization and every iteration; for example, the loop invariant in \e{max} (lines 12 -- 14 in Figure \ref{listing:max}) specifies what the loop has achieved (a part of the final goal) at an intermediate iteration.
\item Loop variant (\e{variant}): an integer measure that remains non-negative and decreases at each loop iteration, ensuring that the loop eventually terminates; the loop variant of the loop in \e{max} is \e{a.count - i} (line 23 in Figure \ref{listing:max}).
\end{itemize}

Contracts embedded in the code allow programmers to open the way to both static analysis of the  program, with AutoProof, and dynamic analysis based on run-time assertion checking, with AutoTest.

\subsection{AutoTest}
\label{autotest}

{AutoTest} \cite{autotest, wei2010satisfying} is an automatic contract-based testing tool: it uses the contracts present in Eiffel programs as {test oracles} and monitors their validity during execution. A test case in AutoTest is an argument-less procedure which calls a ``target routine'' (the routine to be tested) in a certain context (arguments and other objects). The context can be created manually, as in traditional testing environments, but AutoTest can also automatically create it, using sophisticated algorithms. Figure \ref{listing:test_max}  \gt{test\_MAX\_IN\_ARRAY\_max\_1}  shows a test case with target routine  \gt{max} (part of the ``target class'' \e{MAX_IN_ARRAY}).
A test case usually consists of the following components:
\begin{itemize}
\item Declaration of input variables of the {target routine}.
\item Creation of fresh objects for the declared variables.
\item Instantiation of the objects with concrete values.
\item Invocation of the target routine using the instantiated variables.
\end{itemize}

\noindent The execution of a test case in AutoTest performs run-time assertion checking for the {target routine}: AutoTest executes the target routine through a qualified call \footnote{As a reminder, a qualified call is of the form \gt{x.r(args)}, on a target object identified by \gt{x}, as opposed to an unqualified call \gt{r(args)} applying to the current object.} to the routine; during the execution of the routine call, AutoTest evaluates each of the contracts in the routine; if any assertion is violated (evaluated to \e{False}), AutoTest terminates the execution and reports an exception of type ``contract violation''.
The test leads to one of the following verdicts:
\begin{enumerate}
    \item
Passes (the execution of the test case satisfies all specifications).
\item
Fails with a contract violation of the target routine.
\item
Remains unresolved: the violation is not the target routine's fault. The most important case in this category occurs when AutoTest creates an input state that does not satisfy the target routine's precondition. (AutoTest includes optimization strategies to avoid this case, which teaches nothing about the correctness of the software, although it might suggest that the routine's precondition is too strict.)
\end{enumerate}

\noindent AutoTest as described here is a test execution and management framework. AutoTest also includes automatic test case generation mechanisms, not used in the present work.

\subsection{AutoProof}
\label{autoproof}

AutoProof \cite{autoproof, tschannen2015autoproof} checks the correctness of Eiffel programs against their functional specifications (contracts). Verification in AutoProof covers a wide range of properties including whether:
\begin{itemize}
    \item The precondition of a routine holds at the time of call.
    \item The postcondition of a routine holds on routine exit.
    \item The initialization of a loop ensures the invariant.
    \item That initialization leaves the variant non-negative.
    \item The body of a loop preserves the loop invariant (leaves it true if it was true before).
    \item That body leaves the variant non-negative (if it was non-negative before).
    \item It decreases the variant.
    \item A creation procedure (constructor) of a class produces an object that satisfies the class invariant.
    \item A qualified call \e{x.f (...)} finds the target object (the object attached to \e{x}) in a consistent state. AutoProof integrates two mechanisms to verify class invariants: \emph{ownership} \cite{leino2004object} and \emph{semantic collaboration}\cite{polikarpova2014flexible}. A new approach \cite{meyer_class_invariants}, avoiding any need for programmer annotations, is currently being implemented to replace them.
\end{itemize}

\noindent AutoProof relies on Boogie and Z3 as explained next.

\subsection{Boogie and Z3}
\label{boogie_z3}

When verifying an Eiffel program, AutoProof translates it into a Boogie  program, then relies on the Boogie verifier to transform this program into verification conditions for the SMT solver (Z3). The underlying theory is Dijkstra's weakest-precondition calculus\cite{dijkstra1976discipline}.

The verification condition for a routine $r$ of body $b$, precondition $P$ and postcondition $Q$, is of the form

\vspace{0.05in}
  \ \ \ \ \ \ \ \ \ $P$ $\implies$ ($b$ \textbf{wp} $Q$) \hspace{6em} [VC]
\vspace{0.05in}

\noindent where $b$ \textbf{wp} $Q$, per Dijkstra's calculus, is the \textit{weakest precondition} of $b$ for $Q$, meaning the weakest possible assertion such that $b$, started in a state satisfying $P$, will terminate in a state satisfying $Q$. The ``weak/strong'' terminology for logical formulas refers to implication; more precisely, saying that $P'$ is weaker than or equal to $P$ (and $P$ stronger than to equal to $P'$) simply expresses that $P$ $\implies$ $P'$. (Equivalently, if we identify $P$ and $P'$ with the set of states that satisfy the respective assertions, then ``$P$ is stronger than $P'$'' means $P$ $\subset$ $P'$.) Then the verification condition [VC] expresses that the precondition $P$ of the routine is strong enough to guarantee that the routine's execution will yield the postcondition $Q$ (since $P$ is at least as strong as the weakest possible assertion, $b$ \textbf{wp} $Q$, guaranteeing this result).

If property [VC] is satisfied at the Boogie/Z3 level, the Eiffel routine is correct (with respect to $P$ and $Q$).

The technique used by an SMT solver to prove a property such as [VC] is indirect: the solver tries to \textit{satisfy} (the ``S'' in ``SMT'') a given logical formula --- or to determine that such an assignment does not exist. So the formula on which Z3 will work is not [VC] itself but its negation [NVC]:

\vspace{0.05in}
  \ \ \ \ \ \ \ \ \ $\neg$ ($P$ $\implies$ ($b$ \textbf{wp} $Q$)) \hspace{4em} [NVC]
\vspace{0.05in}

\noindent The proof succeeds if \textit{no variable assignment} satisfies [NVC]; in other words, it is impossible to falsify the verification condition.

In the present work, we are interested in the case in which the proof does \textit{not} succeed: the solver does find a variable assignment satisfying [NVC] and hence violating [VC]. The solver has authoritatively  found that the program is buggy, by \textit{disproving} [VC]: it found a set of variable values that causes the routine not to produce a final state satisfying $Q$.

If the goal is not only to give the programmer a simple success/failure notification about the proof, but to help the programmer in the failure case, it is important that such a disproof be \textit{constructive}: it actually identifies a \textit{failing test case}. That test case is not, however, directly usable by the programmer. Instead, it is buried in internal Z3 information and SMT-solving notation. It is the task of Proof2Test to extract the relevant information and turn the counterexample into an actual test case that the programmer can understand, in terms of the original programming language (Eiffel), and run. Section \ref{implementation} will present the details of the counterexample and test generation process in Proof2Test.

Listing \ref{listing:smt encodings} shows how the information appears internally (slightly simplified for the purposes of presentation). The notation is the SMT-LIB format\cite{barrett2010smt}, common to Z3 and other SMT solvers. The goal is to express  [NVC] on line 42. The previous lines define the environment in the form of declarations of types or ``sorts'' (lines 1 -- 2) , functions and constants\footnote{Constants are treated like functions with no arguments, declared, like other functions, in  \gt{declare-fun} clauses.} (3 -- 19), assertions expressing typing properties (20 -- 23) and assertions expressing verification conditions (24 -- 42). The final line, \gt{check-sat}, directs the solver to check satisfaction of the conditions.

\begin{lstlisting}[language = Java,  basicstyle=\fontsize{0.33cm}{0.33cm}, caption = {SMT encodings for verifying \emph{balance\_set}},
    label = {listing:smt encodings}, captionpos=b]
(declare-sort T@U 0)
(declare-sort T@T 0)
(declare-fun type (T@U) T@T)
(declare-fun type_of (T@U) T@U)
(declare-fun intType ( ) T@T)
(declare-fun TypeType ( ) T@T)
(declare-fun FieldType (T@T) T@T)
(declare-fun ACCOUNT ( ) T@U)
(declare-fun balance ( ) T@U)
(declare-fun credit_limit ( ) T@U)
(declare-fun Current ( ) T@U)
(declare-fun amount ( ) Int)
(declare-fun Heap@0 ( ) T@U)
(declare-fun Heap@1 ( ) T@U)
(declare-fun Succ (T@U T@U) Bool)
(declare-fun value ( ) Int)
(declare-fun Select (T@U T@U T@U) T@U)
(declare-fun Store (T@U T@U T@U T@U) T@U)
(declare-fun fun.available_amount (T@U T@U) Int)
(assert (= (type ACCOUNT) TypeType))
(assert (= (type balance) (FieldType intType)))
(assert (= (type credit_limit) (FieldType intType)))
(assert (= (type_of Current) ACCOUNT))
(assert (let (
              (available_amount (fun.available_amount Heap@0 Current))
              (old_balance (Select Heap@0 Current balance))
              (current_balance (Select Heap@1 Current balance))
              (state_transition (Succ Heap@0 Heap@1))
              (update_balance (= Heap@1 (Store Heap@0 Current balance value)))
             )
             (let (
                   (pre (and ($\geq$ amount 0) ($\leq$ amount available_amount)))
                   (wp ($\implies$
                          (= value (+ old_balance amount))
                          (and
                               update_balance state_transition
                               (= current_balance ($-$  old_balance amount))
                           )
                       )
                   )
                 )
                 (not ($\implies$ pre wp))
             )
        )
)
(check-sat)
\end{lstlisting}

\noindent The SMT-LIB syntax is parenthesis-based in the Lisp style, meaning in particular that function application is in prefix form: \gt{(f x)} denotes the result of applying a function \gt{f} to an argument \gt{x}, which in more usual notation would be written \gt{f(x)} . In particular:
\begin{itemize}
    \item
    \gt{($\implies$ x y)} means \gt{x$\implies$y} (\gt{x} implies \gt{y}) in ordinary mathematical notation.
    \item
    \gt{(let a b c)} (to be read as ``let \gt{a} be defined as \gt{b} in \gt{c}'') has the value of \gt{c}, with every occurrence of \gt{a} replaced by \gt{b}.

\end{itemize}

\noindent The final condition \gt{(not ($\implies$ pre wp))} on line 42 (in more usual notation: $\neg$ (\gt{pre} $\implies$ \gt{wp})) corresponds to [NVC], where \gt{wp} is the the weakest precondition. The preceding lines build up auxiliary elements culminating in this definition:

\begin{itemize}

    \item

 Lines 1 -- 2 define the basic \textit{sorts}. ``Sort'' is to the modeling language, SMT-LIB, what ``type'' is to the programming language being modeled; as a result, a sort can describe both values and types from that programming language. Two fundamental sorts used in models generated by Boogie are \gt{T@U} (line 1), describing values in the target programming language, and \gt{T@T}, describing types in that language. The \gt{0} in both declarations indicate the arity (number of arguments, here zero).


    \item

Line 3 specifies that the function \gt{type}, which yields the type of a value, maps a value (\gt{T@U}) to a type (\gt{T@T}).
    \item

The \gt{type\_of} function (line 4) relates an object reference to the object's class. For example: \gt{Current}\footnote{{\ttfamily{Current}} represents the active object in the current execution context, similar to {\ttfamily{this}} in Java.}
is of type \gt{ACCOUNT} (line 23);
\gt{amount}, an integer argument of \e{withdraw}, is declared as an integer constant (line 12); \gt{balance} and \gt{credit\_limit} are two integer fields and defined as \emph{instances} of a composite type \gt{FieldType intType} (lines 21 -- 22).
    \item

Modeling the execution semantics of an object-oriented program requires modeling the behavior of the \emph{heap} (where the objects are stored) through a sequence of constants prefixed with \gt{Heap@}.
In this example, there are two states during execution of \gt{withdraw}: before and after the assignment \e{balance := balance + amount}, represented by \gt{Heap@0} and \gt{Heap@1} (lines 13 -- 14). \gt{Succ} (line 15) specifies the relation between two successive states. The auxiliary variable \gt{value} (line 16) will contain the intermediate result of the assignment --- the result of \e{balance + amount} (line 34).

\item
\gt{Select} and \gt{Store} (lines 17 -- 18) are functions for retrieving and updating the values of data fields.  \gt{Select H obj fd} will give the value in field \gt{fd} of object \gt{obj} in heap \gt{H}. \gt{Store H obj fd new} is a new store obtained by replacing that field value by \gt{new}.
    \item
Line 32 defines the precondition \gt{pre}, rephrased in SMT-LIB from the original Eiffel precondition clause (\e{require}).
\item

Lines 24 -- 30 introduce auxiliary variables.

    \item

Lines 33 -- 40 define \gt{wp}, as the weakest precondition of the routine \e{balance_set} for its given two-clause postcondition (36 -- 37).
\end{itemize}

\section{Proof2Test implementation}
\label{implementation}
Based on the technologies described in the previous section, Proof2Test automatically generates tests from failed proofs. This section presents the overall workflow of Proof2Test, then the details of its implementation.

\subsection{Overview of the Proof2Test process} \label{process}
\begin{figure}[htbp]
\centerline{{\includegraphics[width=3in]{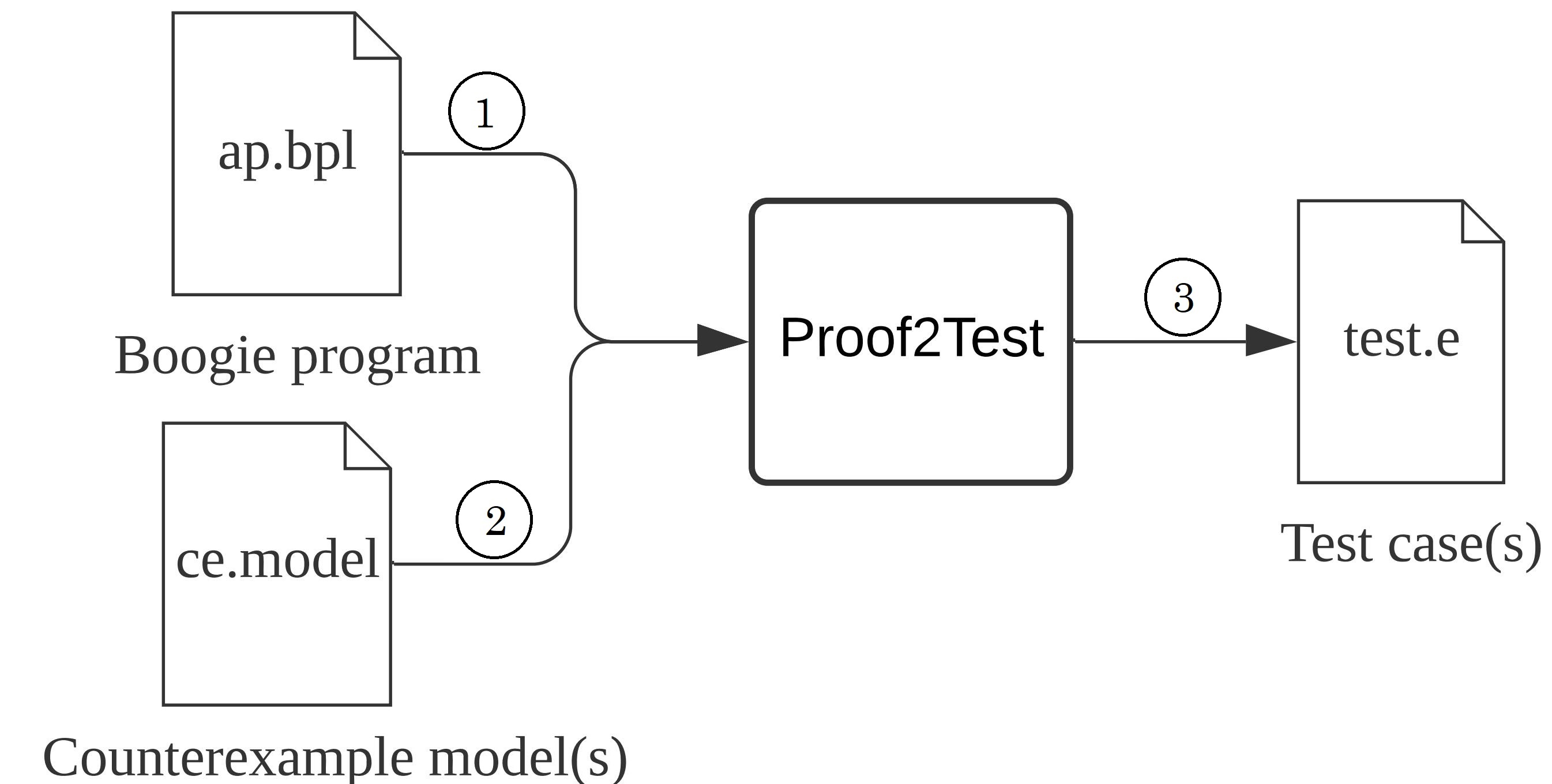}
}}
\caption{Proof2Test workflow}
\label{fig:proof2test-workflow}
\end{figure}

Figure \ref{fig:proof2test-workflow} outlines the workflow of Proof2Test. The inputs are a Boogie program (\gt{ap.bpl}), generated by AutoProof from the input Eiffel program, and an SMT model file  (\gt{ce.model}), containing counterexamples from the Z3 solver. The output is a test script (\gt{test.e}) in the form of an Eiffel class. Proof2Test goes through three steps to construct a test case:
\begin{itemize}[]
    \item \textcircled{{\gt{1}}} Collect the relevant context information from the Boogie program, including names and types of the input arguments of the failed routine $r$, as well as the attributes of the classes involved in $r$.
    \item \textcircled{\gt{2}}  Extract an input vector (a sequence of values of $r$'s input arguments) from a counterexample.
    \item \textcircled{\gt{3}}  Write a test case to the output file \gt{test.e} based on the extracted context information and input vector.
\end{itemize}

\noindent A proof failure in AutoProof corresponds to a contract violation for an erroneous routine of a class, which results in a counterexample.
In the case of multiple proof failures, the resulting model file (\gt{ce.model}) contains the same number of counterexamples. Proof2Test parses each counterexample and generates test cases respectively.



A \textbf{counterexample} is an execution trace (a sequence of program states) of the erroneous routine $r$, at the end of which the program reaches a failed state (violating a contract element of $r$).
Proof2Test parses the model to obtain the trace's input vector, from which it produces a test case for AutoTest (Section \ref{autotest}).

\subsection{Extraction of input vector from counterexample}
In the example of Figure \ref{listing:ACCOUNT}, the routine \e{withdraw} is incorrect since its body does not ensure the postcondition \textit{balance\_set}  (line 42). AutoProof consequently fails to prove the routine correct by obtaining, through Boogie, a counterexample from Z3. Figure \ref{listing:counterexample} shows that counterexample (key parts only, slightly simplified).
\begin{figure}[htbp]
    \centering
\begin{minipage}{.3\textwidth}
\begin{lstlisting}[language = Java, basicstyle=\fontsize{0.33cm}{0.33cm}]
amount -> 1
Current -> T@U!val!18
Heap@0 -> T@U!val!17
Heap@1 -> T@U!val!22
ACCOUNT -> T@U!val!6
balance -> T@U!val!7
credit_limit -> T@U!val!8
value -> 11
type_of -> {
  T@U!val!18 -> T@U!val!6 }
\end{lstlisting}
\end{minipage}
\begin{minipage}{.60\textwidth}
\begin{lstlisting}[firstnumber=last, language = Java, basicstyle=\fontsize{0.33cm}{0.33cm}]
Succ -> {
    T@U!val!17  T@U!val!22 -> True }
Store -> {
    T@U!val!17   T@U!val!18   T@U!val!7   11 -> T@U!val!22 }
Select -> {
    T@U!val!17   T@U!val!18   T@U!val!7 -> 10
    T@U!val!17   T@U!val!18   T@U!val!8 -> ($-$ 20)
    T@U!val!22   T@U!val!18   T@U!val!7 -> 11 }
fun.available_amount -> {
    T@U!val!17  T@U!val!18 -> 30 }
\end{lstlisting}
\end{minipage}
    \caption{Counterexample of the proof failure of \emph{balance\_set}}
    \label{listing:counterexample}
\end{figure}

\noindent

\noindent In this example the path to a contract violation goes through only two states, called \gt{Heap@0} and \gt{Heap@1} (as defined in lines 3 and 4). The execution trace is not given explicitly but can be inferred from the differences between \gt{Heap@0} and \gt{Heap@1}. Formally, the counterexample in Figure \ref{listing:counterexample} is a sequence of definitions of name-value associations, in the form  \gt{name} $\rightarrow$ \gt{value}. Basic values, other than integers, are of the form \gt{T@U!val!n} denoting an abstract location \gt{n}. For example, the value of \gt{amount} in the counterexample is 1 (line 1) and \gt{balance} is stored at abstract location \gt{7}.

The initial and final heaps, \gt{Heap@0} and \gt{Heap@1}, are associated (lines 3 and 4) with locations \gt{17} and \gt{22}. To infer the execution trace, it suffices to look at the definitions starting with line 11, which show the differences between the contents of these two heaps by referring to \gt{T@U!val!17} and \gt{T@U!val!22}.

The preceding lines define variable values (for example, \gt{balance}, \gt{credit\_limit}).

For a function \e{f} of $n$ arguments, the specification of the function's value in the counterexample takes the form
\begin{itemize} []
\item \gt{a$_1$} \gt{a$_2$} ... \gt{a$_n$} $\rightarrow$ \gt{x}
\item \gt{b$_1$} \gt{b$_2$} ... \gt{b$_n$} $\rightarrow$ \gt{y}
\item ...
\end{itemize}
\noindent to mean that  \gt {f(}\gt{a$_1$,} \gt{a$_2$,} ... \gt{a$_n$} \gt{)} = \gt{x} etc. For example lines 15 to 18 give the state of both the initial heap (\gt{T@U!val!17}) and the final heap (\gt{T@U!val!22}) by specifying the \gt{Select} function. (As seen in Section \ref{boogie_z3}, \gt{Select (H, obj, fd)} is the value in field \gt{fd} of object \gt{obj} in heap \gt{H}.) This specification yields the values, in either or both heaps, of the
\gt{balance} and \gt{credit\_limit} fields (\gt{T@U!val!7} and \gt{T@U!val!8}) for the \e{Current} object (\gt{T@U!val!18}). Lines 13 and 14 define a state change, in the form of an update of the \gt{Store} function: change the \gt{balance} field (\gt{T@U!val!7}) of the current object (\gt{T@U!val!18}) in \gt{Heap@0} so that it will have the value \gt{11} in \gt{Heap@1}. In programming language terms this would be written just \e{balance := 11} (where \e{balance} denotes \e{Current.balance}).

To construct the input vector of \e{withdraw} from the model, it suffices to obtain the following information:
\begin{enumerate}
\item The initial state of the \e{Current} object (the target object of the qualified call), which includes the initial values of the two data fields \gt{balance} and \gt{credit\_limit}.
\item  The value of the only argument \e{amount}.
\end{enumerate}

\noindent On can obtain the value of \e{amount} directly (line 1), and the values of \e{balance} and \e{credit\_limit} from  their symbolic addresses and the \gt{Select} function (16 -- 17). The resulting input vector for \e{withdraw} is:  \e{Current.balance} = 10, \e{Current.credit\_limit} = $-$ 20,  \e{amount}  = 1.

\subsection{Construction of test cases based on input vectors}
Once it has obtained the input vector, Proof2Test starts constructing a test case in the form of a qualified call, here   \gt{current\_object.withdraw(amount)}. The input vector contains the values of \gt{current\_object} and  argument \gt{amount} before the call.

\begin{figure}[!htbp]
\begin{lstlisting}[basicstyle=\fontsize{0.3cm}{0.3cm}, captionpos=b]
test_ACCOUNT_withdraw_1
	local
        current_object: ACCOUNT; amount: INTEGER
	do
		  create current_object.make(0)
		  {INTERNAL}.set_integer_field (balance, current_object, 10)
		  {INTERNAL}.set_integer_field (credit_limit, current_object, (- 20))
		  amount := 1
		  current_object.withdraw (amount)
	end
\end{lstlisting}
\caption{Test case generated from the proof failure of \e{withdraw}}
\label{listing:test_withdraw}
\end{figure}

\noindent Figure \ref{listing:test_withdraw} shows the generated test case, which includes four components:
\begin{enumerate}
\item Local declaration of the target object (\e{current\_object}) and input argument (\gt{amount}) (line 3).
\item Creation of the target object (line 5).
\item Initialization of the target object and argument (lines 6 -- 8). Since Eiffel is a strongly typed language, setting individual fields of arbitrary objects requires using low-level mechanisms from the library class \e{INTERNAL}, such as \e{set_integer_field}.
\item Qualified call to \e{withdraw} (line 9).
\end{enumerate}

\begin{algorithm}[hbt!]
\caption{Generate a single test case based on data collected from a counterexample} \label{alg:test}
\begin{algorithmic}[1]
\item \textbf{Input:} target class \gt{c}, target routine \gt{r}, input vector \gt{v}
\item \textbf{Output:} a test case \gt{t}
\State \gt{t} := \gt{name\_of\_test\_case(c, r)}  \hspace*{\fill}{\gt{-}\gt{-} Construct the name for the test case}
\State \gt{t\_d} := \gt{declare\_variable(current\_object)}
\State \gt{t\_c} := \gt{create\_reference(current\_object)}
 \hspace*{\fill}{\gt{-}\gt{-} Declare and create  \gt{current\_object}}
\State \gt{t\_i} := \gt{empty\_string}

\State \textbf{across} {input arguments of \gt{r} \textbf{as} \gt{a}} \textbf{loop}
     \State \ \ \ \ \gt{t\_d} := \gt{t\_d} $+$ \gt{declare\_variable(a)}
      \hspace*{\fill}{\gt{-}\gt{-} Declare each argument}
     \State \ \ \ \ \textbf{if} {\gt{a} is a variable of primitive type} \textbf{then}
     \State \ \ \ \ \ \ \ \ \ \ \gt{value} := \gt{get\_value(a, v)}
      \hspace*{\fill}{\gt{-}\gt{-} Get the value of \gt{a} from \gt{v}}
     \State \ \ \ \ \ \ \ \ \ \ \gt{t\_i} := \gt{t\_i} $+$ \gt{assignment(a, value)}
 \hspace*{\fill}{\gt{-}\gt{-} Use assignments to instantiate primitive-type arguments}
     \State \ \ \ \ \textbf{else}
     \State  \ \ \ \ \ \ \ \ \ \
     \gt{t\_c} := \gt{t\_c} $+$ \gt{creation\_of\_reference(a)}
      \hspace*{\fill}{\gt{-}\gt{-} Create objects for reference arguments}
     \State \ \ \ \ \ \ \ \ \ \  \gt{t\_i} := \gt{t\_i} $+$ \gt{instantiate\_reference(a, v)}
      \hspace*{\fill}{\gt{-}\gt{-} Instantiate reference arguments according to \gt{v}}
     \State \ \ \ \ \textbf{end}
     \State \textbf{end}
     \State \textbf{if} {\gt{r} is a function} \textbf{then}
     \State \ \ \ \ \gt{res} := \gt{name\_of\_result\_variable(r)}
     \State \ \ \ \ \gt{t\_d} := \gt{t\_d} $+$
     \gt{declare\_variable(res)}
      \hspace*{\fill}{\gt{-}\gt{-} Construct a variable \gt{res} to fetch \gt{r}'s result}
    \State \textbf{end}
    \State \gt{t} := \gt{t} $+$ \gt{t\_d} $+$ \gt{t\_c} $+$ \gt{t\_i}
    \hspace*{\fill}{\gt{-}\gt{-}  Combine the clauses of declaration, creation and instantiation}
    \State \gt{t} := \gt{t} $+$ \gt{call\_routine(current\_object, r, res)}
     \hspace*{\fill}{\gt{-}\gt{-} Call \gt{r} with its arguments}
\end{algorithmic}
\end{algorithm}

\noindent The test generation process applies Algorithm \ref{alg:test}. The result of the algorithm is the code of a test case \gt{t}, which includes three components: declaration clauses (\gt{t\_d}), creation clauses (\gt{t\_c}) and instructions instantiating variables (\gt{t\_i}).

The algorithm loops over all the input arguments (lines 7 -- 15) of \gt{r} and instantiates each of them according to their types: arguments of primitive types (\e{INTEGER}, \e{BOOLEAN}, \e{CHARACTER} and \e{REAL}) are instantiated through direct assignments (line 10) to the corresponding values in the input vector (for example, the integer argument \e{amount} is instantiated as \e{amount := 1}); arguments of reference types are instantiated by applying the \gt{instantiate\_reference} procedure (see the details in Algorithm \ref{alg:instantiate}) based on the input vector. Finally,  the algorithm builds a call of the target routine (lines 16 -- 21) with the instantiated arguments.

\begin{algorithm}[hbt!]
\caption{\gt{instantiate\_reference}: instantiate a variable of reference type} \label{alg:instantiate}
\begin{algorithmic}[1]
\item \textbf{Input:} object reference \gt{o}, input vector \gt{v}
\item \textbf{Output:} text \gt{t}
\State \textbf{if} {\gt{o} has an instantiated alias} \textbf{then}
    \State \ \ \ \ \gt{alias} := \gt{get\_alias\_of\_reference(o)}
    \State \ \ \ \ \gt{t} := \gt{t} $+$ \gt{assignment(o, alias)}  \hspace*{\fill}{\gt{-}\gt{-}  Construct an assignment \gt{o := alias}}
    \State \textbf{elseif} {\gt{o} is a variable of container type} \textbf{then}
         \State \ \ \ \ \gt{size} := \gt{get\_size\_of\_container(o, v)}  \hspace*{\fill}{\gt{-}\gt{-} Get \gt{o}'s size}
         \State \ \ \ \ \gt{items} := \gt{get\_item\_of\_container(o, v)}  \hspace*{\fill}{\gt{-}\gt{-} Get \gt{o}'s elements}
         \State \ \ \ \ \textbf{from}
         \State  \ \ \ \ \ \ \ \ \ \ \gt{i} := 1
         \State \ \ \ \ \textbf{until}
         \State \ \ \ \ \ \ \ \ \ \ {\gt{i} $\leq$ \gt{size} }
         \State \ \ \ \ \textbf{loop}
              \State \ \ \ \ \ \ \ \ \ \ \gt{t} := \gt{t} $+$ \gt{force\_clause(o, items[i])}
               \hspace*{\fill}{\gt{-}\gt{-}  Use \gt{force} routine to insert each of \gt{o}'s elements}
              \State \ \ \ \ \ \ \ \ \ \ \gt{i} := \gt{i} + 1
              \State \ \ \ \ \textbf{end}
      \State \textbf{else}
	    \State \ \ \ \ \textbf{across} {each field of \gt{o} \textbf{as} \gt{f}} \textbf{loop}
           \State \ \ \ \ \ \ \ \ \ \ \textbf{if} {\gt{f} is a variable of primitive type} \textbf{then}
			    \State \ \ \ \ \ \ \ \ \ \ \ \ \ \ \ \  \gt{value} := \gt{get\_value(f, v)}
			    \State \ \ \ \ \ \ \ \ \ \ \ \ \ \ \ \  \gt{t} := \gt{set\_field\_clause(f, value)}
			     \hspace*{\fill}{\gt{-}\gt{-}  Use \gt{set\_type\_field} function to instantiate \gt{f}}
			\State  \ \ \ \ \ \ \ \ \ \ \textbf{else}
			    \State \ \ \ \ \ \ \ \ \ \ \ \ \ \ \ \ \gt{t} :=  \gt{instantiate\_reference(f, v)}
			\State \ \ \ \ \ \ \ \ \ \ \textbf{end}
		\State \ \ \ \  \textbf{end}
	 \State \textbf{end}
    \end{algorithmic}
\end{algorithm}

Algorithm \ref{alg:instantiate} instantiates an object reference \gt{o} based on the input vector \gt{v}:
\begin{itemize}
    \item
Checks whether \gt{o} has an alias that has been instantiated earlier.
\item
If so, construct a direct assignment to the earliest alias (lines 4 -- 5).
\item
If \gt{o} is an object of a container type (such as \e{ARRAY} and \e{SEQUENCE}), instantiates each of its fields (lines 6 -- 16).
\item
If it is instead an object of a non-container reference type, instantiate its fields transitively: assign to  fields of primitive types (such as \e{balance} and \e{credit\_limit}) their corresponding values in the input vector (lines 20 -- 21); for fields of reference types, apply the procedure recursively (line 23).
\end{itemize}

\section{Combining proofs and tests: the process}
\label{diagnosis}

As noted in the introduction, tests and proofs have often been considered distinct, incompatible or competing approaches to software verification. The assumption behind the present work is that it is better to view them as complementary.

\subsection{General description} \label{overall-process}

\begin{figure}[htbp]
\centerline{{\includegraphics[width=6.7in]{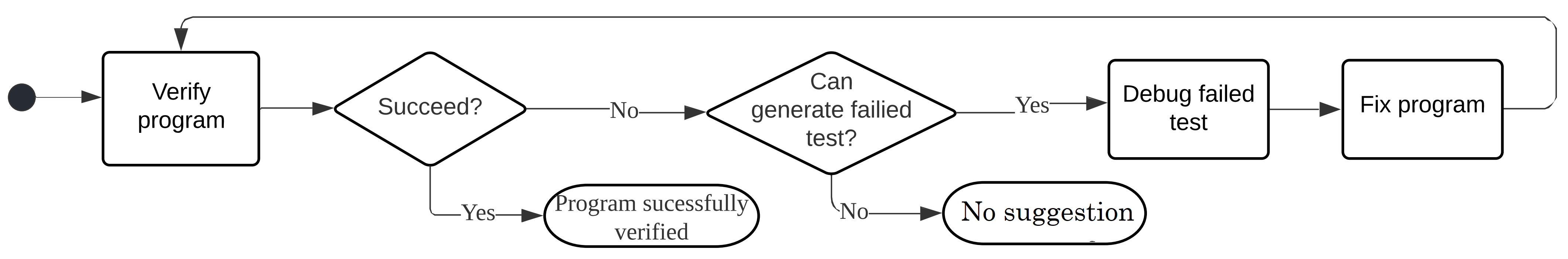}
}}
\caption{Program verification with the assistance of Proof2Test.}
\label{fig:workflow}
\hfil
\end{figure}

\begin{figure}[htbp]
\centerline{{\includegraphics[width=5.8in]{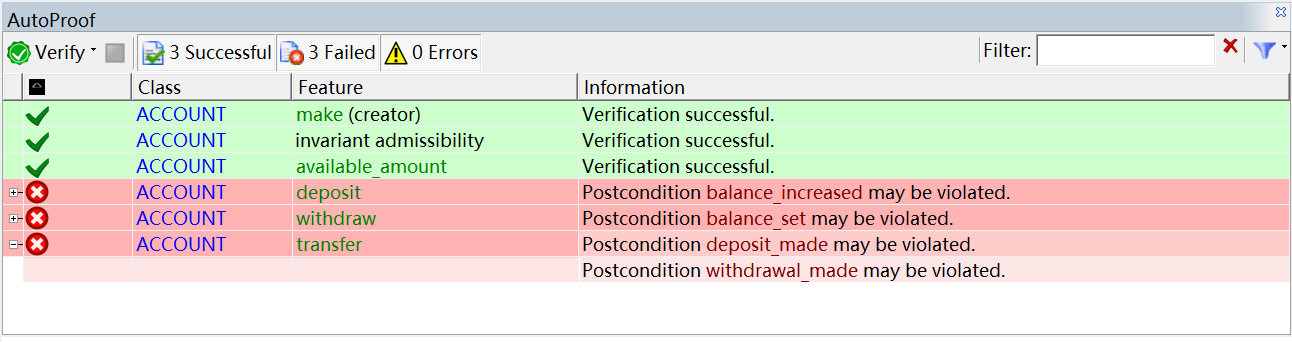}
}}
\caption{Verification result of \e{ACCOUNT} in AutoProof, with some properties proved and others not}
\label{fig:verification}
\hfil
\end{figure}

The verification process combining both approaches, thanks to AutoProof, AutoTest and Proof2Test, is the following, illustrated by Figure \ref{fig:workflow} :

\begin{itemize}
    \item
Step A:
Attempt the verification with AutoProof.
If the verification succeeds, the process stops. The remaining steps assume the verification failed.
Figure \ref{fig:verification} shows an example in which AutoProof is not able to prove some properties (although it does succeed with some others).
\item
Step B:
Run Proof2Test. The remaining steps assume Proof2Test is able to generate a failing test (if not, Proof2Test is not helpful in this case).
\item
Step C:
Run the test.
\item
Step D: use the test to attempt to correct the bug using standard testing and debugging techniques. The fix can be a change to the code or to the contract (or occasionally both).
\item
Repeat the process, attempting to prove the corrected code, until the proof succeeds.
\end{itemize}

When Proof2Test is not able to generate a failing tests (the workflow in Figure \ref{fig:workflow} terminates with  ``No suggestion''), there are two possible reasons:
\begin{itemize}
    \item the generated tests are passed, mostly due to the weakness of the specifications (loop invariants or postconditions of a supplier routine) of the target routine; Section \ref{experiment} contains a more detailed discussion of such cases.
    \item the program contains the data types or program constructs that are not supported by Proof2Test; in this case, the test generation either terminates in an abnormal state (indicating a limitation of Proof2Test) or terminates normally but the resulting tests would be incorrect and thus not compilable.
\end{itemize}


\subsection{Running the verification process: an example}
\label{verification-example}

Here is the application of the above process to the
 \e{ACCOUNT} class from Section \ref{eiffel}.

 Step A: run AutoProof on the class;
Figure \ref{fig:verification} already showed the result, which displays AutoProof's inability to establish four postconditions: \emph{balance\_increased} in the \e{deposit} routine, \emph{balance\_set} in \e{withdraw}, \emph{withdrawal\_made} and \emph{deposit\_made} in \e{transfer}.
In this case, the model file (\gt{ce.model}) contains four counterexample models, each of which corresponds to a proof failure (postcondition violation).

\begin{figure}[htbp]
\begin{lstlisting}[basicstyle=\fontsize{0.3cm}{0.3cm}, captionpos=b]
test_ACCOUNT_deposit_1
    local
        current_object: ACCOUNT; amount: INTEGER
    do
        create current_object.make (0)
        {INTERNAL}.set_integer_field (balance, current_object, 38)
        {INTERNAL}.set_integer_field (credit_limit, current_object, (-62))
        amount := (-5)
        current_object.deposit (amount)
    end

test_ACCOUNT_withdraw_1
	local
        current_object: ACCOUNT; amount: INTEGER
	do
		  create current_object.make (0)
		  {INTERNAL}.set_integer_field (balance, current_object, 10)
		  {INTERNAL}.set_integer_field (credit_limit, current_object, (- 20))
		  amount := 1
		  current_object.withdraw (amount)
	end
	
test_ACCOUNT_transfer_1
    local
        current_object, other: ACCOUNT;  amount: INTEGER
    do
        create current_object.make (0)
        create other.make (0)
        {INTERNAL}.set_integer_field (balance, current_object, (-2))
        {INTERNAL}.set_integer_field (credit_limit, current_object, (-32))
        amount := 6
        {INTERNAL}.set_integer_field (balance, other, (-83))
        {INTERNAL}.set_integer_field (credit_limit, other, (-83))
        current_object.transfer (amount, other)
    end

test_ACCOUNT_transfer_2
    local
        current_object, other: ACCOUNT; amount: INTEGER
    do
        create current_object.make (0)
        {INTERNAL}.set_integer_field (balance, current_object, (-2))
        {INTERNAL}.set_integer_field (credit_limit, current_object, (-53))
        amount := 1
        other := current_object
        current_object.transfer (amount, other)
    end
\end{lstlisting}
\caption{Test cases for \e{ACCOUNT}}
\label{listing:test_account}
\end{figure}

Step B: use Proof2Test to generate test cases, shown in Figure \ref{listing:test_account}, from the four counterexamples. In the order of proof failures:
\begin{itemize}
\item \e{test\_ACCOUNT\_deposit\_1} (lines 1 -- 10) corresponds to \e{deposit}'s postcondition \emph{balance\_increased}; it first instantiates \e{current\_object} by setting the values of the \e{balance} and \e{credit\_limit} to 38 and $-$62 (lines 6 -- 7) and then sets the value of the input argument \e{amount} to $-5$ (line 8); finally it invokes a qualified call of \e{deposit} (the target routine) with the instantiated argument (line 9).

\item \e{test\_ACCOUNT\_withdraw\_1} (lines 12 -- 21) corresponds to the proof failure of \e{withdraw}; similar to the test case of \e{deposit}, it instantiates \e{current\_object}'s two fields, \e{balance} and \e{credit\_limit}, to 10 and $-$20, and sets the value of \e{amount} to $1$, followed by a call to \e{withdraw}.

\item \e{test\_ACCOUNT\_transfer\_1} (lines 23 -- 35) corresponds to the postcondition \emph{deposit\_made} of \e{transfer}; it first instantiates the state of \e{current\_object} (lines 29 -- 30), and then initializes the two arguments of \e{transfer}, \e{amount} (line 31) and \e{other} (line 32 -- 33); it then calls \e{transfer} using the two arguments.

\item \e{test\_ACCOUNT\_transfer\_2} (lines 37 -- 47) corresponds to the proof failure of postcondition \emph{withdrawal\_made} of \e{transfer}; it follows the same structure of \e{test\_ACCOUNT\_transfer\_1}: it starts with the initialization of \e{current\_object} and arguments, followed by a call to \e{transfer};
additionally, as \e{other} and \e{current\_object} are aliases (they have the same symbolic value in the counterexample model), instead of instantiating \e{other} using the \e{set\_type\_field} routines, it directly assign \e{other} with \e{current\_object} (line 45).
\end{itemize}

\begin{figure}[htbp]
\centerline{{\includegraphics[width=5.6in]{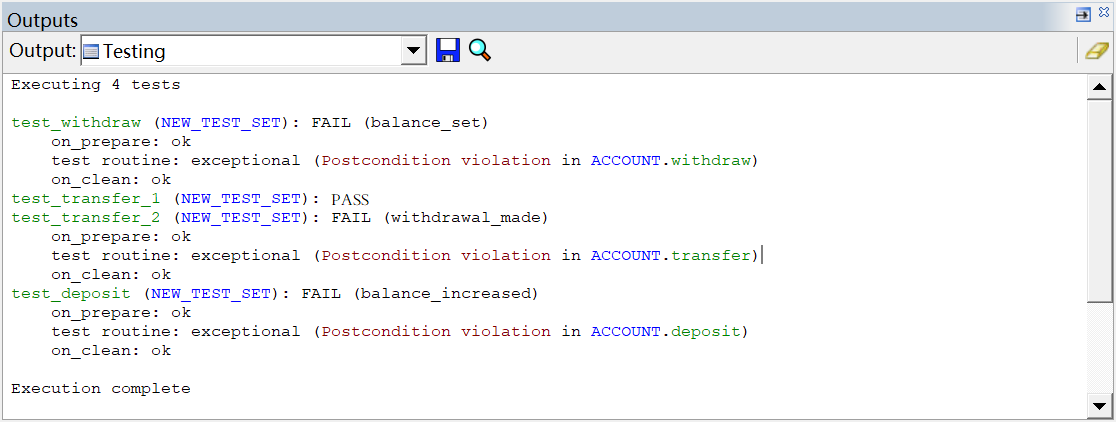}
}}
\caption{Testing results in AutoTest: ``PASS'' means running the test satisfies all assertions during execution; ``FAIL'' indicates the occurrence of a contract violation in the test run.}
\label{fig:testing}
\hfil
\end{figure}

Step C: exercise the test cases in AutoTest. Figure \ref{fig:testing} shows the testing result: among the four test cases, \e{test\_ACCOUNT\_transfer\_1} passed and the other three test cases failed, raising the same postcondition violations as in the verification result (Figure \ref{fig:verification}).

\begin{figure}
  \centering
  \subfigure[Execution trace of {\gt{deposit}}]{
  \includegraphics[width=3.35in]{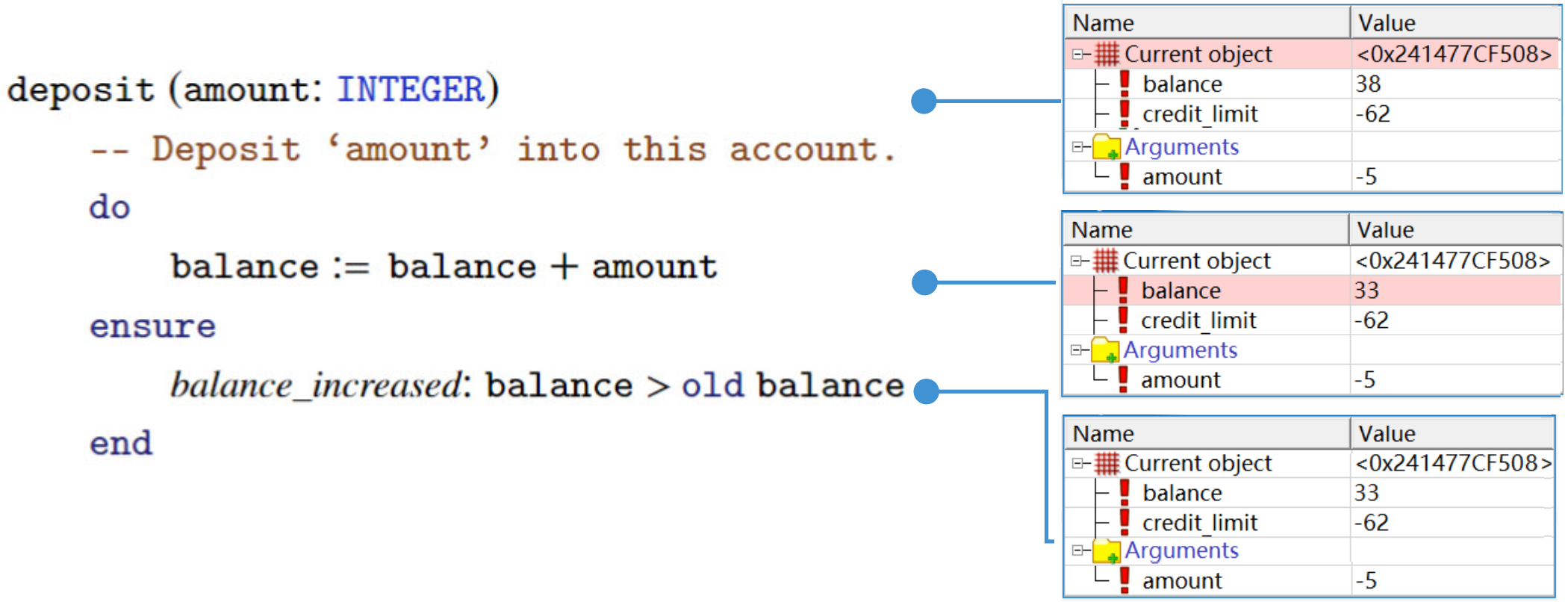}}
  \subfigure[Execution trace of \gt{withdraw}]{
  \includegraphics[width=3.55in]{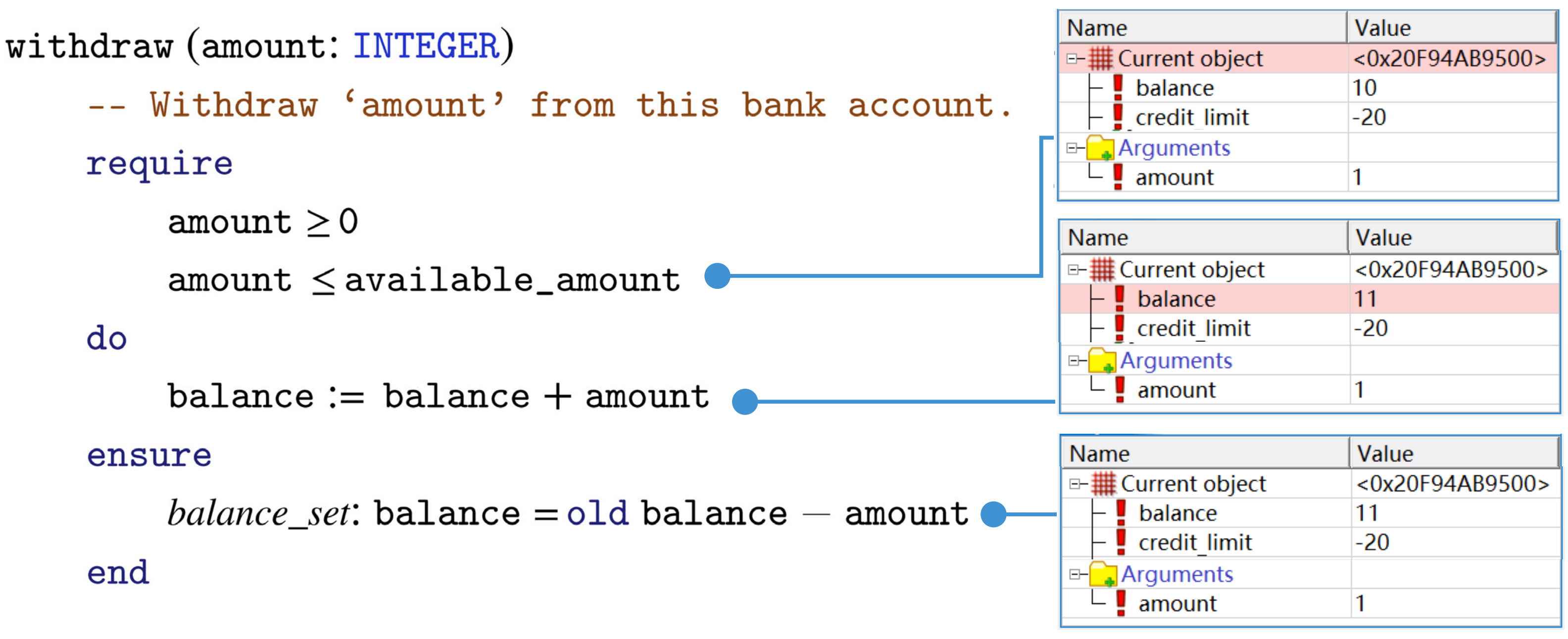}}
    \subfigure[Execution trace of \gt{transfer}]{
  \includegraphics[width=5.3in]{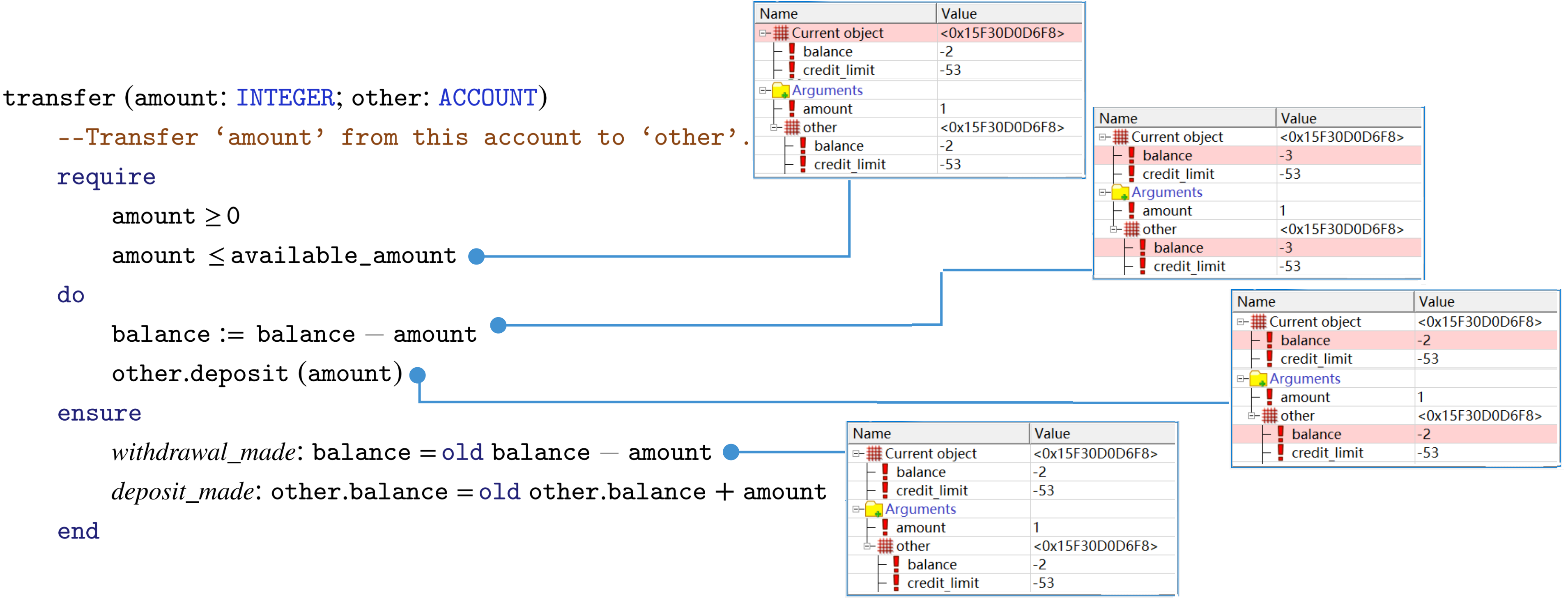}}
  \caption{Debugging failed tests in EiffelStudio: each block displays a program state (a list of variables and their values) after executing a designated statement (starting point of a blue line); the variable will be highlighted in red when its value is changed after executing the statement.}
  \label{fig:debug}
\end{figure}

Step D: to understand the causes of the contract violations, use the debugger from EiffelStudio to step through the three failed test cases: \e{test\_ACCOUNT\_deposit\_1}, \e{test\_ACCOUNT\_withdraw\_1} and \e{test\_ACCOUNT\_transfer\_2}.
For each test case,  move forward to the line of the respective call of the target routine and advance the execution of the target routine on one instruction at a time to observe how the program state (values of variables) evolves. Figure \ref{fig:debug} shows the execution traces of the three target routines:
\begin{itemize}
\item \e{deposit} (Figure \ref{fig:debug} (a)): initially the \gt{balance} of the \gt{Current} object is 38; after execution of the assignment, \gt{balance} is set to 33 (the sum of its previous value and \gt{amount}; the final state violates the postcondition \emph{balance\_increased} as the value of \gt{balance} (33) is smaller than its old value (38). The execution trace reveals a flaw of \e{deposit}:
depositing negative \e{amount} is permitted. To correct this error, a precondition should be added to require that the value of the input argument \e{amount} should be non-negative.

\item \e{withdraw} (Figure \ref{fig:debug} (b)): at the initial state, \e{amount} is 1, \e{balance} is 10, \e{credit_limit} is $-$ 20, \e{available_amount}  (the difference between \e{balance} and \e{credit limit}) is 30; the two preconditions are satisfied; after executing the assignment, \e{balance} is set to 11 (the sum of its previous value and \e{amount}); at the final state, the value of \e{old balance} - \e{amount} (9) is different with the value of \e{balance} (11), which violates the postcondition. This indicates the apparent bug: the routine implementation (addition operation) is inconsistent with what is expected in the postcondition (subtraction operation). Replacing the ``+'' with ``$-$'' is enough to correct the bug.

\item \e{transfer} (Figure \ref{fig:debug} (c)): initially,  \e{amount} = 1 and \e{available_amount} is 51, which satisfies the two preconditions; after withdrawing \e{amount} from the \e{Current} account (\e{balance := balance $-$ amount}), \e{balance} of the \e{Current} account is set to $-$3; moreover, since \e{Current} and \e{other} are aliases (they refer to the same object), \e{balance} of the \e{other} account is also set to $-$3; similarly, after depositing \e{amount} to the \e{other}, the \gt{balance} in both \e{other} account and \e{Current} account are updated to $-$2; at the end state, the postcondition \emph{withdral\_made} is not satisfied. The execution trace shows a erroneous scenario where the \e{Current} account is transferring money to itself. To exclude this particular case from the execution, we can add a precondition \e{Current /= other} to limit that \e{other} should be an account other than \e{Current}.
\end{itemize}

\noindent These failed tests become part of the regression test suite of the \emph{target class} \e{ACCOUNT}.






\section{Experiment and evaluation}
\label{experiment}
A preliminary evaluation of the usability of Proof2Test used the example programs of Table \ref{table: experiment}. Altogether, the experiment involved 49 editions of 9 Eiffel programs. The report includes not only the test results but also the test generation times for both the proofs and the tests. 

The process is the following:

\begin{itemize}
\item
Each edition results from injecting a single error into a correct program. The faults, injected manually, include:
\begin{itemize}[]
    \item 1. Switching between $+$ and $-$, $\leq$ and $<$, $>$ and $\geq$.
    \item 2. Removing a clause from a loop invariant, a precondition or a postcondition.
    \item 3. Removing a statement in a routine body.
    \item 4. Swapping two subsequent statements in a routine body.
\end{itemize}
In general, the decisions of which instruction or other clause to manipulate and how to mutate operators is random, with the goal of covering  as many candidate areas as possible and creating faults that are similar to those made by programmers in practice.

\item
When verifying the modified program, AutoProof fails on a single assertion of one of the following seven kinds: postcondition violation at the end of a call; class invariant violation; precondition violation on entry to a call; loop invariant violated after loop initialization; loop invariant not maintained by loop body;  loop variant not decreased; loop variant becoming negative.

\item
From each such failure, Proof2Test generates one test case.
\end{itemize}

\begin{table*}[htbp]
 \scriptsize
  \centering
   \renewcommand\arraystretch{1.3}
  \caption{Test generation from failed proofs on a collection of examples}
    \begin{tabular}{|p{70pt}|p{15pt}|p{60pt}|p{84pt}|p{88pt}|p{17pt}|p{30pt}|p{32pt}|p{16pt}|}
    \hline
    Class  & Size$_P$ & Routine & Violated assertion & Failure type & Size$_M$ & Proof time (s) & Test gen. time (s) & Test result
    \\ \hline
    {ACCOUNT} & 97 & withdraw & balance\_set & postcondition violation & 658 & 0.247 & 0.241 & Fail \\ \cline{4-9}
    & &  & balance\_non\_negative & class invariant violation & 620 & 0.275 & 0.225 & Fail \\ \cline{3-9}
    &   & deposit & balance\_set & postcondition violation & 654 & 0.241 & 0.202 & Fail
    \\ \cline{3-9}
     & & transfer & amount $\leq$ 10 & precondition violation & 633 & 0.248 & 0.212 & Fail \\ \cline{4-9}
     & &  & withdrawal\_made & postcondition violation &  638 & 0.243 & 0.208 & Fail
     \\ \cline{4-9}
     & &  & withdrawal\_made & postcondition violation & 672 & 0.253 & 0.214 & Pass
    \\ \cline{4-9}
     & &  & deposit\_made & postcondition violation & 660 & 0.244 & 0.223 & Pass
    \\ \hline
    \multirow{2}{*}{CLOCK} & 131 & increase\_hours & valid\_hours & precondition violation  & 572 & 0.253 & 0.245 & Fail  \\ \cline{3-9}
    & & increase\_minutes & hours\_increased & postcondition violation  & 607 & 0.251 & 0.231
    & Fail  \\ \cline{4-9}
    & & & hours\_increased & postcondition violation  & 641 & 0.263 & 0.218 & Pass
    \\ \cline{4-9}
    & & & minutes\_increased & postcondition violation  & 606 & 0.259 & 0.222 & Pass
    \\ \cline{3-9}
    & & increase\_seconds & valid\_seconds & precondition violation & 574 & 0.241 & 0.201 & Fail
    \\ \cline{4-9}
    & & & hours\_increased & postcondition violation & 647 & 0.243 & 0.219 & Pass
    \\ \cline{4-9}
    & &  & minutes\_increased & postcondition violation & 625 & 0.248 & 0.201 & Pass
    \\ \cline{4-9}
    & & & valid\_minutes & precondition violation & 577 & 0.245 & 0.192 & Fail
    \\ \hline
    {HEATER} & 73 & turn\_on\_off & heater\_remains\_off & postcondition violation & 633 & 0.615 & 0.207 & Fail
    \\ \cline{4-9}
    &  & & heater\_remains\_on & postcondition violation & 643 & 0.642 & 0.213 & Fail
    \\ \cline{4-9}
    &  &  & heater\_remains\_on & postcondition violation & 631 & 0.250 & 0.212 & Fail
    \\ \cline{4-9}
    &  &  & heater\_remains\_off & postcondition violation & 633 & 0.246 & 0.181 & Fail
    \\ \hline
    {LAMP} & 71 & turn\_on\_off & turn\_off & postcondition violation & 728 & 0.250 & 0.194 & Fail \\ \cline{4-9}
    & & & turn\_off & postcondition violation & 696 & 0.278 & 0.205 & Fail \\ \cline{3-9}
      & & adjust\_light & from\_high\_to\_low & postcondition violation & 632 & 0.265 & 0.222 & Fail
      \\ \cline{4-9}
      & &  & from\_medium\_to\_high & postcondition violation  & 624 & 0.270 & 0.208 & Fail
    \\ \hline
    \multirow{2}{*}{BINARY\_SEARCH} & 50 & binary\_search & present & postcondition violation  & 1243 & 0.327 & 0.471 & Fail  \\ \cline{4-9}
    &  &  & not\_in\_lower\_part & invariant not maintained  & 1312 & 0.39 & 0.71 & Fail
    \\ \cline{4-9}
    &  &  & not\_in\_lower\_part & invariant not maintained & 1376 & 0.395 &0.440 & Fail
    \\ \cline{4-9}
    &  &  & -- & variant not decreased & 1370 & 0.325 & 0.398 & Fail
    \\ \cline{4-9}
    &  &  & not\_in\_upper\_part & invariant not maintained & 1120 & 0.288 & 0.323 & Fail
    \\ \cline{4-9}
    &  &  & present & postcondition violation & 1347 & 0.550 & 0.382 & Pass
    \\ \cline{4-9}
    &  &  & not\_in\_lower\_part & invariant violated on entry & 1227 & 0.356 & 0.366 & Fail
    \\ \hline
    {LINEAR\_SEARCH} & 27 & linear\_search & result\_in\_bound & invariant violated on entry & 955 & 0.709 & 0.332  & Fail
     \\ \cline{4-9}
     & & & present & postcondition violation  & 977 & 0.283 & 0.373 & Fail
     \\ \cline{4-9}
     & & & present & postcondition violation  & 975 & 0.280 & 0.344 & Pass
          \\ \cline{4-9}
     & & & -- & variant being negative & 976 & 0.279 & 0.312 & Fail
    \\ \hline
      {MAX\_IN\_ARRAY} & 33 & max\_in\_array & max\_so\_far & invariant not maintained& 1104 & 0.288 &0.282 & Fail
     \\ \cline{4-9}
     & & & is\_maximum & postcondition violation  & 1059 & 0.284 & 0.313 & Pass
     \\ \cline{4-9}
     & & & result\_in\_array & postcondition violation  & 1049 & 0.278 & 0.307 & Pass
     \\ \cline{4-9}
     & & & is\_maximum & postcondition violation  & 1057 & 0.286 & 0.323 & Fail
     \\ \cline{4-9}
     & & & max\_so\_far & invariant violated on entry & 992 & 0.290 & 0.303 & Fail
     \\ \cline{4-9}
     & & & i\_in\_bounds & invariant violated on entry & 1015 & 0.282 & 0.315 & Fail
    \\ \hline
    {SQUARE\_ROOT} & 38 & square\_root & -- & variant not decreased & 648 & 0.258 & 0.192 & Fail
     \\ \cline{4-9}
     & & & valid\_result & postcondition violation & 723 & 0.257 & 0.205 & Pass
     \\ \cline{4-9}
     & & & valid\_result & invariant not maintained  & 644 & 0.308 & 0.181 & Fail
          \\ \cline{4-9}
     & & & result\_so\_far & invariant not maintained  & 631 & 0.400 & 0.194 & Fail
     \\ \hline
      {SUM\_AND\_MAX} & 36 & sum\_and\_max & is\_maximum & postcondition violation & 1402 & 0.301 & 0.336 & Pass
       \\ \cline{4-9}
      & &  & is\_maximum & postcondition violation & 1405 & 0.321 & 0.333 & Fail
      \\ \cline{4-9}
      & &  & partial\_sum\_and\_max & invariant not maintained & 1178 & 0.307 & 0.333 & Fail
      \\ \cline{4-9}
       & &  & partial\_sum\_and\_max & invariant violated on entry & 1127 & 0.743 & 0.346 & Fail
      \\ \cline{4-9}
      & &  & sum\_max\_non\_negative & invariant not maintained & 1178 & 0.800 & 0.313  & Fail
      \\ \cline{4-9}
      & &  & sum\_in\_range & postcondition violation & 1375 & 0.310 & 0.362  & Fail
    \\ \hline
     {Total} & 556 & 14 & 49 & & 44069 & 15.968 & 13.985 & 37/12
    \\ \hline
    \end{tabular}%
  \label{table: experiment}%
\end{table*}%

\noindent Each row in Table \ref{table: experiment} corresponds to an experimental task performed on a program variant, including three procedures:
1) verification of the program in AutoProof;
2) generation of test case in Proof2Test;
3) exercising the test case in AutoTest.
The Size$_P$ and Size$_M$ columns provide the number of lines of the program and counterexample model.
The ``Proof time'' column lists the time in seconds for obtaining a proof in AutoProof, and the ``Test generation time''  column  the time Proof2Test takes to generate the test from the corresponding model. The last column shows the results of running the corresponding tests in AutoTest.

The examples include: 
1) the class \e{ACCOUNT} introduced in Section \ref{eiffel};
2) a class \e{CLOCK} implementing a clock counting seconds, minutes, and hours;
3) a \e{HEATER} class implementing a heater, adjusting it state (on or off) based on the current temperature and user-defined temperature;
4) a class \e{LAMP} describing a lamp equipped with a switch (for switching the lamp on or off) and a dimmer (for adjusting the light intensity of the lamp);
5) a class \e{BINARY\_SEARCH}  implementing binary search;
6) a class \e{LINEAR\_SEARCH} implementing linear search;
7) the class \e{MAX_IN_ARRAY} presented in Section \ref{example};
8) a class \e{SQUARE\_ROOT} for calculating two approximate square roots of a positive integer;
9) a class \e{SUM\_AND\_MAX}  computing the maximum and the sum of the elements in an array.
\hyperref[results for account]{A.1} and \hyperref[results for linear search]{A.2} detail the experiment results of \e{ACCOUNT} and \e{LINEAR_SEARCH}; the results for other examples can be found in the technical report \cite{testdetails}.

Among the 49 total test runs of the generated counterexamples, 37 failed but 12 runs passed. Ideally, we would like all counterexample to fail execution; the discrepancy reflects a difference of the semantics for proofs and tests. If a routine \e{r} calls a routine \e{s}:
\begin{itemize}
    \item The proof of \e{r} relies only on the specification of \e{s}, independently of its implementation. In other words, proofs use modular semantics.
    \item A test of \e{r}, in particular an execution with run-time assertion monitoring under EiffelStudio and AutoTest, must execute \e{r} and hence must call \e{s} as it is, relying on its implementation. In other words, tests use global (non-modular) semantics.
\end{itemize}

\noindent As a consequence, assuming the prover (AutoProof) is sound, a test failure implies a proof failure but the reverse is not necessarily true: the counterexample generated by a failed proof (based on modular semantics) might not lead to a failure when we execute it (under non-modular semantics).

When such discrepancies arise in practice, they are typically due to a routine \e{s} with an implementation that is ``correct'' in some intuitive sense (it does what the programmer informally intended) but a specification that is not complete: it does not express all the relevant properties of the routine, which the prover would need to prove the correctness of the calling routine \e{r}.

\noindent The passing counterexample test runs are therefore useful, as the failing ones are, but in a different way: they usually alert the developers to the presence of a specification in need of improvement. (This case is the most common one. There always remains in principle the possibility of a mere limitation of the prover which -- because of fundamental undecidability results --- cannot be both sound and complete. However tempting it may be to blame the prover for proof failures, in practice this theoretical limitation seldom hits with a powerful prover such as the Boogie/Z3 combination for AutoProof: the culprit is almost always the programmer, not the tool.)


Incomplete specifications are indeed the reason for the 12 proof failures in the experiment. For example, failures in Variant \hyperref[account: variant 6]{6} and \hyperref[account: variant 7]{7} of \e{ACCOUNT} are due to weakness of the postconditions of the associated calling routines, and the failure in the Variant \hyperref[linear search: variant 3]{3} of \e{LINEAR_SEARCH} to incomplete loop invariants.

For the 37 failures for which the tests fail, the causes can be categorized as follows:
\begin{itemize}
\item Postcondition violation due to incorrect implementations (examples in this category include the Variant \hyperref[account: variant 1]{1}, \hyperref[account: variant 4]{4}, \hyperref[account: variant 5]{5} of \e{ACCOUNT}, and the Variant \hyperref[linear search: variant 2]{2} of \e{LINEAR_SEARCH}).
\item Precondition violation due to the inconsistency of specification between a routine and its calling routine; for example, the failure in Variant \hyperref[account: variant 3]{3} of \e{ACCOUNT} is caused by the inconsistency of specification between the routine  \e{transfer} and its calling routine \e{deposit}.
\item Precondition violation of a calling routine due to a fault in its client routine; for example, the verification of Variant {1} of \e{CLOCK} (for the details of this variant, see the technical report \cite{testdetails}) results in a failure of the precondition of a routine \e{set_hours} when calling it from another routine \e{increase_hours}.
\item Violation of loop invariant at start of loop due to an incorrect implementation of loop initialization (see Variant \hyperref[linear search: variant 1]{1} of \e{LINEAR_SEARCH} as an example).
\item Loop invariant not maintained due to incorrect exit condition or faults in the loop body; examples in this category include Variant 4 of \e{BINARY_SEARCH} and Variant 1 of \e{MAX_IN_ARRAY}.
\item Loop variant not decreased due to incorrect exit condition of the loop; an example in this category is Variant 1 of \e{SQUARE_ROOT}.
\item Loop variant be negative due to the incorrectness of the loop variant; an example is the Variant \hyperref[linear search: variant 4]{4} of \e{LINEAR_SEARCH}).
\end{itemize}

\noindent In most cases, the failing test is useful: executing it yields a specific trace illustrating how the program leads to the same contract violation that makes the proof fail.
Even in cases such as Variant \hyperref[account: variant 2]{2} and \hyperref[account: variant 5]{5} of \e{ACCOUNT} in which the test input itself immediately demonstrates the problem, executing the automatically generated to allows programmers to step into the debugger and understand the issue in depth by going into step-by-step mode.

In all cases, the generated tests are important as regression tests: once the corresponding bug has been corrected, every test should become part of the project's regression suite, an essential tool for further development of the project.

\section{Related Work}
\label{related_work}

The idea of using counterexamples to generate test cases is not new, but it has been mostly applied to verification approaches using model checking of temporal-logic specifications \cite{beyer2004generating, fantechi2005enhancing, black2000modeling, beyer2018tests}. We are only aware of one existing attempt \cite{nilizadeh2022generating} (building on work on using counterexamples for automatic program repair\cite{nilizadeh2021more}) to apply the idea in the context of Hoare-style verification; it exploits counterexamples produced by OpenJML \cite{cok2021jml} (a verification tool for Java programs) to generate unit tests in JUnit \cite{cheon2002simple} format.
The tool described in that article needs both to generate counter-examples from an SMT server and to generate, from the JML source,  Java code instrumented to monitor some of the assertions (expressed as comments in the original code) at run-time. 
It then uses the concrete tests, extracted from counterexamples, to call the code. Proof2Test does not need any extra code generation since it uses inputs from counterexamples to call the erroneous routines, taking advantage of the built-in runtime assertion checking mechanism of EiffelStudio. We benefit here from the integration of contracts as a basic feature of the language and environment, rather than an add-on to a non-contract language such as Java.

KeyTestGen \cite{engel2007generating, ahrendt2016proof} is a test generation tool based on the KeY, which is an automatic proof system for Java programs \cite{ahrendt2014key} based on dynamic logic. The KeyTestGen makes use of the branch information (in the form of proof trees), generated during program proofs, to construct path constraints with respect to different branches. It then applies an SMT solver to generate test data that satisfy those path constraints. As a result, the produced tests cases cover different program branches (paths). In contrast, Proof2Test aims to produce tests that can reproduce the proof failures at run time; it directly exploits counterexamples that are available in cases of proof failures and thus requires no extra SMT solving process.

In line with the objective of helping verification engineers understand the reasons of proof failures, many approaches have been proposed to provide a more user-friendly visualization of counterexample models: Claire et al. \cite{le2011boogie} developed the Boogie Verification Debugger (BVD), which can interpret a counterexample model as a static execution trace (a sequence of abstract states). BVD has been applied in Dafny \cite{leino2010dafny} and VCC \cite{cohen2009vcc}. David et al. \cite{hauzar2016counterexamples} transformed the models back into a counterexample trace comprehensible at the original source code level (SPARK) and display the trace using comments. Similarly, Stoll \cite{stoll2019smt} implemented a tool that translates the models into programs understandable at the Viper source code level \cite{muller2016viper}.
Chakarov et al. \cite{chakarov2022better} transformed SMT models to a format close to the Dafny syntax.
Instead of providing a static diagnosis trace, the Proof2Test approach is able to present a ``dynamic’’ trace, through execution of tests produced based on the models, which allows verification engineers to produce a program trace leading to a failing run-time state. Verification engineers can even use a debugger to step through the test, observing the faulty behaviors at their own pace. We find this approach more appropriate since every software developer is used to working with a debugger.


Another way of facilitating proof failures diagnosis is to generate of useful counterexamples: Polikarpova et al. \cite{polikarpova2013run} developed a tool Boogaloo, which applied symbolic execution to generate counterexamples for failed proofs of Boogie programs. Boogaloo displays the resulting counterexamples in the form of valuations of relevant variables, similar to the intermediate result of the present work after the step ``extraction of input data from the model’’ in Section \ref{implementation}.  Likewise, Petiot et al. \cite{petiot2018testing, petiot2016your} developed STADY that produces failing tests for the failed assertions using symbolic execution techniques. Their approach is also referred to testing-based counterexample synthesis: they first translated the original C program into programs suitable for testing (run-time assertion checking), and then applied symbolic execution to generate counterexamples (input of the failing tests) based on the translated program. Unlike that approach, Proof2Test exploits the original programs, that are inherently amenable to run-time assertion checking and thus requires no additional program transformation. Furthermore, our approach directly makes use of the counterexample models produced by the underlying provers and no extra counterexample generation process is performed. However, their approach is more fine-grained than ours as they can distinguish between specification weaknesses failures and prover limitations failures.

Other facilities for diagnosing proof failures: M\"{u}ller et al. \cite{muller2011using} implemented a Visual Studio dynamic debugger plug-in for Spec\#, to reproduce a failing execution from the viewpoint of the prover. This approach creates a variation of the original program for debugging, based on a modular verification semantics: the effect of the iteration of a loop is represented by the corresponding values in the counterexample. Tschannen et al. \cite{tschannen2013program} proposed a ``two-step'' approach to narrow down the reasons of proof failures: it compares the proof failures with those of its variant where called functions are inlined and loops are unrolled, which allows to discern failures caused by specification weaknesses and failures resulting from inconsistency between code and contract. However, inlining and unrolling are  limited to a given number of nested calls and explicit iterations.

Some recent approaches share the present work's vision of combining static and dynamic techniques to make verification more usable. Julian et al. \cite{tschannen2011usable} proposed a combination of  AutoProof and AutoTest at a higher level, where the two tools were integrated in the Eiffel IDE to avail the complementarity of proof and testing. On one hand, for those programming features currently not supported by AutoProof (such as the code that relies on external precompiled C functions), testing can be used to check the correctness of routines. On the other hand, proof can be used to analyze the code that can not be tested (e.g., deferred functions that have no implementations).
Collaborative verification \cite{christakis2012collaborative} is also based on the combination of testing and static verification, and on the explicit formalization of the restrictions of each tool used in the combination.
The Proof2Test approach also complements the limitations of proof techniques with testing, with a particular purpose to improve user experience in understanding proof failures. However, it does not integrate the results of the two different techniques.

\section{Limitations} \label{limitations}

The results obtained in this study rely on a specific combination of technologies (section \ref{technology_stack}): Eiffel as the programming language, contract-equipped programs, and the AutoProof tool stack relying on Boogie, itself based on the Z3 SMT solver.

That last brick is the easiest to replace since Proof2Test relies not on the specifics of Z3 but on its SMT-LIB interface, which current SMT solvers generally share.

More generally, the setup assumes a Hoare-style verification framework (of which Boogie is but one example), and a language that supports the corresponding constructs (preconditions, postconditions, class invariants). Examples of such languages include the JML (Java Modeling Language) \cite{leavens1998jml} extension of Java, the Spark \cite{carre1990spark} extension of Ada, and the Spec\# \cite{barnett2005spec}  extension of C\#. We have not studied the possible application of the ideas to completely different verification frameworks, based for example on abstract interpretation or model checking.

The current version of Proof2Test is subject to the following limitations:

\begin{itemize}
    \item It does not support the more advanced parts of the Eiffel system, in particular generic classes.Data structures are limited to arrays and sequences.
    \item It generates tests for individual routines (methods). There is no mechanism at this point to generate tests for an entire program.
\end{itemize}

These limitations will need to be removed for Proof2Test to be applicable to industrial-grade programs.

\section{Conclusion and future work}
\label{conclusion}

The key assumption behind this work is that program proofs (static) and program tests (dynamic) are complementary rather than exclusive approaches. Software verification is hard; we should take advantage of all techniques that help. Proofs bring the absolute certainties that tests lack, but are abstract and hard to get right; tests cannot guarantee correctness but, when they fail, bring the concreteness of counterexamples, immediately understandable to the programmer and opening up the possibility of using well-understood debugging tools. Relying on the seamless integration of AutoProof and AutoTest in the Eiffel method and the EiffelStudio environment, Proof2Test attempts to leverage the benefits of both.

From this basis, work is proceeding in various directions:

\begin{itemize}

\item
Cover the language constructs and types that (as noted in section \ref{limitations}) are not yet handled.

    \item
Generate failing tests when the tests from Proof2Test are successful or non-executable.

\item
Make Proof2Test (currently a separate tool)  a part of the EiffelStudio tool suite, at the same level of integration as AutoProof and AutoTest.

\item
Extend the scope of the work to include support for more intricate specifications, such as class invariants and assertions involving ghost states.
\item
On the theoretical side, develop a detailed classification of proof failures and possible fix actions with correspondence to their categories.
\item
On the empirical side, perform a systematic study of the benefits of the proposed techniques for verification non-experts.

\end{itemize}

\noindent As it stands, we believe that Proof2Test advances the prospect of an effective approach to software verification combining the power of modern proving and testing techniques.

\vspace{0.4cm}
\noindent \textit{Acknowledgments} Alexander Kogtenkov and Alexandr Naumchev made important contributions to the discussions leading to this article. We are grateful to Filipp Mikoian and Manuel Oriol for useful discussions. We thank Amirfarhad Nilizadeh, one of the authors of the pioneering counterexample-based JML-related work cited earlier \cite{nilizadeh2022generating}, for useful comments on an early version. We are indebted to the reviewers of that early version for insightful comments that led to significant improvements of the article.

\bibliography{reference}

\newpage
\section*{Appendix: detailed results for selected programs}
\label{appendix}
The detailed results and analysis of applying Proof2Test to the sample classes listed in the article appear in a technical report \cite{testdetails}. The material in this Appendix, extracted from that report, covers classes \e{ACCOUNT} and \e{LINEAR_SEARCH}.

All runs took place on a Windows 11 machine with a 2.1 GHz Intel 12-Core processor
and 32 GB of memory. AutoProof or Proof2Test was the only computationally-intensive
process running during the experiments. Version numbers for the underlying techology are:
EiffelStudio 22.05; Boogie 2.11.1.0; Z3 4.8.14. On average, AutoProof
ran for 0.326 seconds for each program; Proof2Test ran for 0.285 each test generation,
with each test generation run producing one test case, from a single
counterexample model.

\subsection*{A.1 Results for class \e{ACCOUNT}}
\label{results for account}
\begin{figure}[htbp]
\centerline{{\includegraphics[width=4.6in]{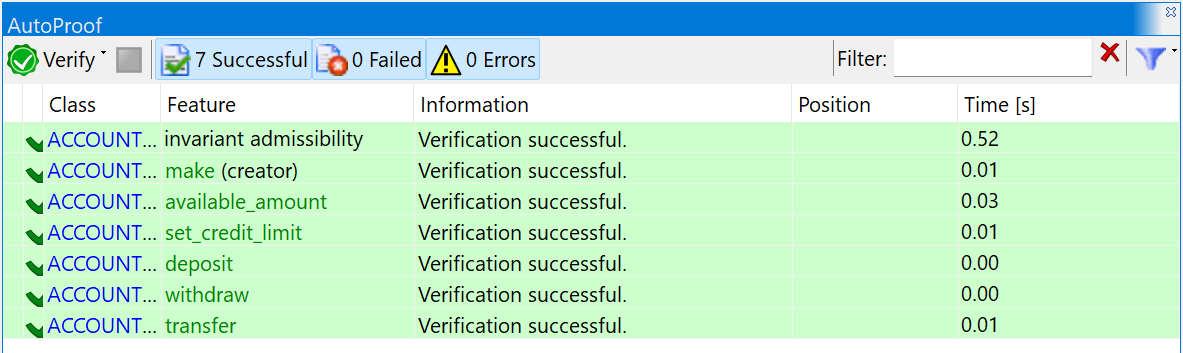}}}
\caption{Verification result of \e{ACCOUNT} in AutoProof: all routines are verified successfully (highlighted with green), which indicates that implementations of those routines are correct with respect to their specifications.}
\label{fig: account_proof_result}
\end{figure}

Below shows a correct version of the \e{ACCOUNT} class, which includes a set of features representing basic operations on bank account: \e{deposit} (line 51), \e{withdraw} (line 64), and \e{transfer} (line 77). Figure \ref{fig: account_proof_result} displays its verification result, which suggests a complete functional correctness. To demonstrate how Proof2Test can generate tests from proof failures, different faults are introduced into the correct version, resulting in 5 faulty variants.

\begin{lstlisting}[captionpos=b, basicstyle=\fontsize{0.27cm}{0.27cm}]
class
    ACCOUNT

create
    make

feature {NONE} -- Initialization
    make
        -- Initialize empty account.
        note
            status: creator
        do
            balance := 0
            credit_limit := 0
        ensure
            balance_set: balance = 0
            credit_limit_set: credit_limit = 0
        end

feature -- Access

    balance: INTEGER
        -- Balance of this account.

    credit_limit: INTEGER
        -- Credit limit of this account.

    available_amount: INTEGER
        -- Amount available on this account.
        note
            status: functional
        do
            Result := balance - credit_limit
        end

feature -- Basic operations

    set_credit_limit (limit: INTEGER)
        -- Set `credit_limit' to `limit'.
        require
            limit_not_positive: limit <= 0
            limit_valid: limit <= balance
        do
            credit_limit := limit
        ensure
            modify_field ([``credit_limit'', ``closed''], Current)
            credit_limit_set: credit_limit = limit
        end


    deposit (amount: INTEGER)
        -- Deposit `amount' in this account.
        require
            amount >= 0
        do
            balance := balance + amount
        ensure
            modify_field ([``balance'', ``closed''], Current)
            balance_increased: balance >= old balance
            balance_set: balance = old balance + amount
        end


    withdraw (amount: INTEGER)
        -- Withdraw `amount' from this account.
        require
            amount_not_negative: amount >= 0
            amount_available: amount <= available_amount
        do
            balance := balance - amount
        ensure
            modify_field ([``balance'', ``closed''], Current)
            balance_set: balance = old balance - amount
            balance_decrease: balance <= old balance
        end

    transfer (amount: INTEGER; other: ACCOUNT_1)
        -- Transfer `amount' from this account to `other'.
        note
            explicit: wrapping
        require
            amount_not_negative: amount >= 0
            amount_available: amount <= available_amount
            other /= Current
        do
            withdraw (amount)
            other.deposit (amount)
        ensure
            modify_field ([``balance'', ``closed''], [Current, other])
            withdrawal_made: balance = old balance - amount
            deposit_made: other.balance = old other.balance + amount
        end

invariant
    credit_limit_not_positive: credit_limit <= 0
    balance_non_negative: balance - credit_limit >= 0
end

\end{lstlisting}

\begin{figure}
  \centering
  \subfigure[]{
  \includegraphics[width=4.8in]{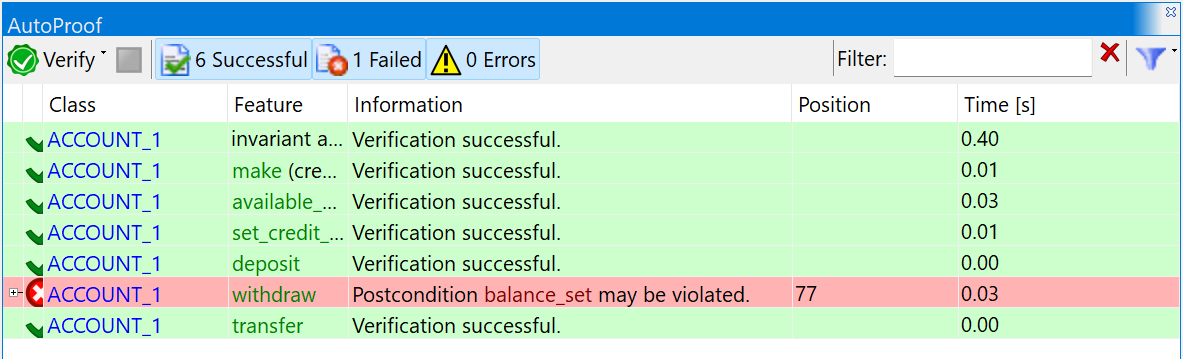}
  \label{fig: account_1_proof_result}
  }
  \subfigure[]{
  \includegraphics[width=4.4in]{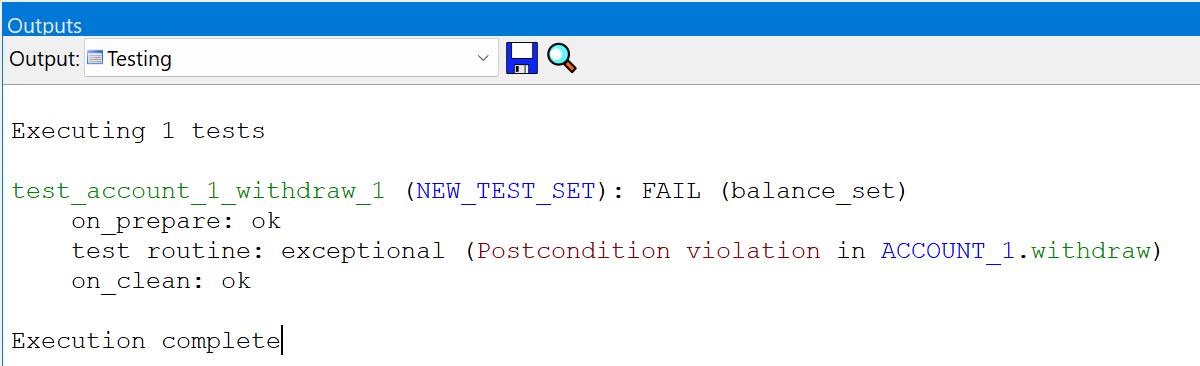}
  \label{fig: account_1_test_result}}
  \caption{(a) Verification result of \e{ACCOUNT_1} in AutoProof; (b) Testing result of \e{test_ACCOUNT_1_withdraw_1} in AutoTest}
  \label{fig:account_1}
\end{figure}

\noindent \textbf{Variant 1 of \e{ACCOUNT}}
\label{account: variant 1}
\begin{itemize}[$\bullet$]
  \item {Fault injection}: at line 73, change the postcondition \e{balance_set} from ``\e{balance = old} \e{balance - amount}'' into ``\e{balance = old} \e{balance + amount}''.
  \item {Resulting failure}: as shown in Figure  \ref{fig: account_1_proof_result}, the fault results in a violation of postcondition \e{balance\_set} of the \e{withdraw} procedure.
  \item {Cause of the failure}: the implementation of \e{withdraw} (which \emph{deduces} \e{balance} by \e{amount}) and specification (which requires the \emph{increment} of \e{balance} by \e{amount}) is inconsistent.
  \item Proof time: 0.247 sec
  \item Test generation time: 0.241 sec
  \item Resulting test case: Figure  \ref{fig: account_1_test_case} shows the test case generated from the failure --- it calls \e{withdraw} with input \e{balance = 11797}, \e{credit_limit = -1}, and \e{amount = 11798};
  \item Testings result: running the test case, as shown in Figure  \ref{fig: account_1_test_result}, raises an exception of violation of \e{balance_set}, which maps to the same failure in AutoProof.
  \item Comment: the value of the test input does not contain specific meaning to the failure; as the failure is caused by inconsistency between implementation and specification, executing the \e{withdraw} procedure with any valid test input (which satisfies the precondition) would raise the same contract violation as in the proof.
\end{itemize}

\begin{figure}
\centering
\begin{lstlisting}[captionpos=b, basicstyle=\fontsize{0.27cm}{0.27cm}]
	test_ACCOUNT_1_withdraw_1
		local
			current_object: ACCOUNT_1
			amount: INTEGER_32
		do
			create current_object.make
			{P_INTERNAL}.set_integer_field_ (``balance'', current_object, 11797)
				-- current_object.balance = 11797
			{P_INTERNAL}.set_integer_field_ (``credit_limit'', current_object, (-1))
				-- current_object.credit_limit = (-1)
			amount := 11798
			current_object.withdraw (amount)
		end
\end{lstlisting}
\caption{Test from the failed proof of \e{balance_set}}
\label{fig: account_1_test_case}
\end{figure}

\begin{figure}
  \centering
  \subfigure[]{
  \includegraphics[width=5.2in]{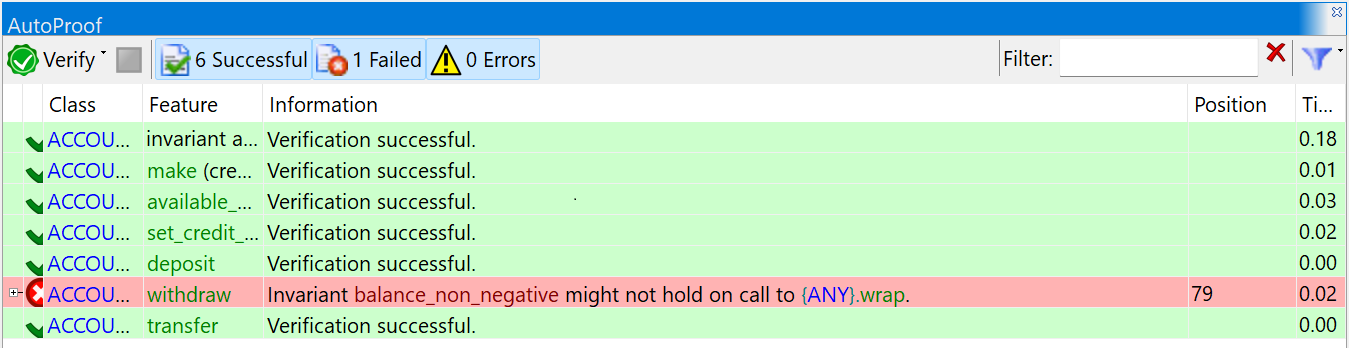}
  \label{fig: account_2_proof_result}
  }
  \subfigure[]{
  \includegraphics[width=4.2in]{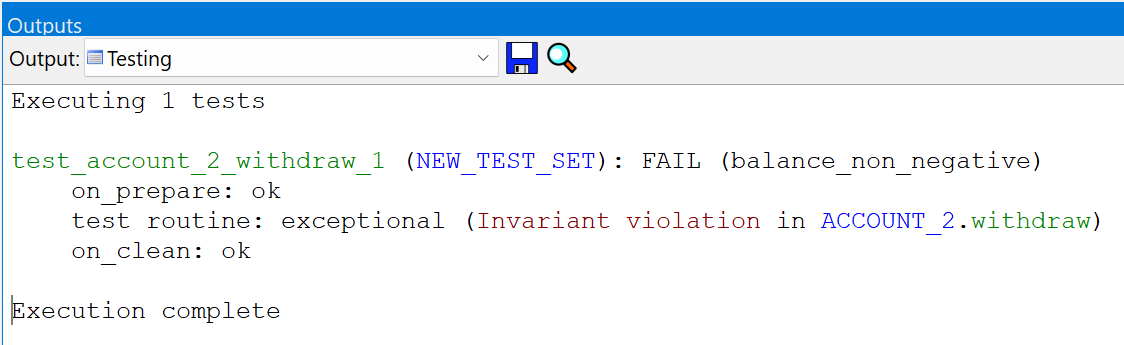}
  \label{fig: account_2_test_result}}
  \caption{(a) Verification result of \e{ACCOUNT_2} in AutoProof; (b) Testing result of \e{test_ACCOUNT_2_withdraw_1} in AutoTest}
  \label{fig:account_2}
\end{figure}

\newpage
\noindent \textbf{Variant 2 of \e{ACCOUNT}}
\label{account: variant 2}
\begin{itemize}[$\bullet$]
  \item {Fault injection}: at line 68, remove the precondition \e{amount_available} of \e{withdraw}.
  \item {Resulting failure}: as shown in Figure  \ref{fig: account_2_proof_result}, the class invariant \e{balance\_non\_negative} (line 96), which states that the balance (represented by \e{balance - amount}) should not be negative, is violated. (Note that a class invariant which is supposed to hold at the entry and exit of every routine.)
  \item {Cause of the failure}: the precondition of \e{withdraw} is too weak; there should be a precondition to constrain the amount permitted in a withdrawal operation.
  \item Proof time: 0.275 sec
  \item Test generation time: 0.225 sec
  \item Resulting test case: Figure  \ref{fig: account_2_test_case} shows the test case from Proof2Test, which calls \e{withdraw} with input --- \e{balance = 0}, \e{credit_limit = 0}, \e{amount = 1}.
  \item Testings result: running the test case in raises the failure of invariant \e{balance_non_negative} (same failed contract in the proof) as shown in Figure  \ref{fig: account_2_test_result}.
  \item Comment: the test input itself immediately demonstrates the problem -- when there is no money left in the account, withdrawal operation should be forbidden; the executable test, however, is still useful in this case: programmers can still choose run the test and switch to the debugging mode to see how it fails step by step; the test can also be a part of the test suite for regression testing in later development stages.
\end{itemize}

\begin{figure}
\centering
\begin{lstlisting}[captionpos=b, basicstyle=\fontsize{0.27cm}{0.27cm}]
		test_ACCOUNT_2_withdraw_1
		local
			current_object: ACCOUNT_2
			amount: INTEGER_32
		do
			create current_object.make
			{P_INTERNAL}.set_integer_field_ (``balance'', current_object, 0)
				-- current_object.balance = 0
			{P_INTERNAL}.set_integer_field_ (``credit_limit'', current_object, 0)
				-- current_object.credit_limit = 0
			amount := 1
			current_object.withdraw (amount)
		end
\end{lstlisting}
\caption{Test case from failed proof of \e{balance_non_negative}}
\label{fig: account_2_test_case}
\end{figure}

\begin{figure}
  \centering
  \subfigure[]{
  \includegraphics[width=5.2in]{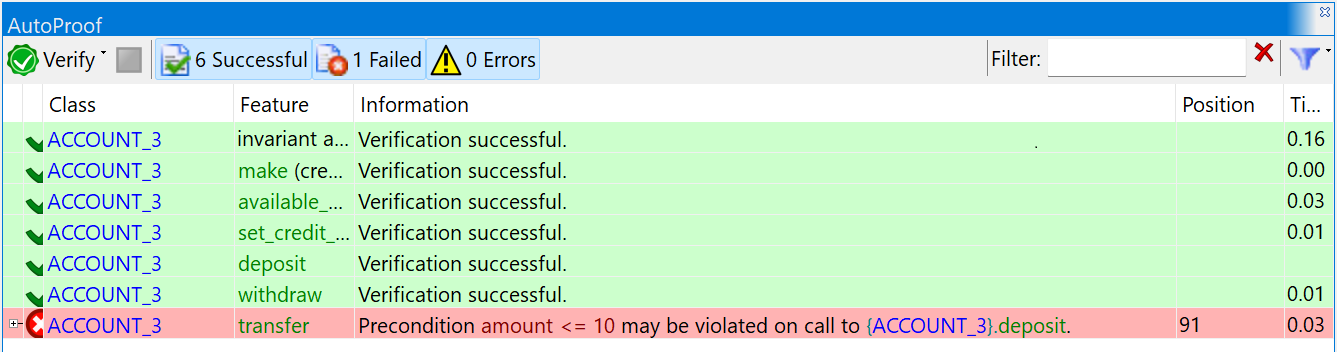}
  \label{fig: account_3_proof_result}
  }
  \subfigure[]{
  \includegraphics[width=4.2in]{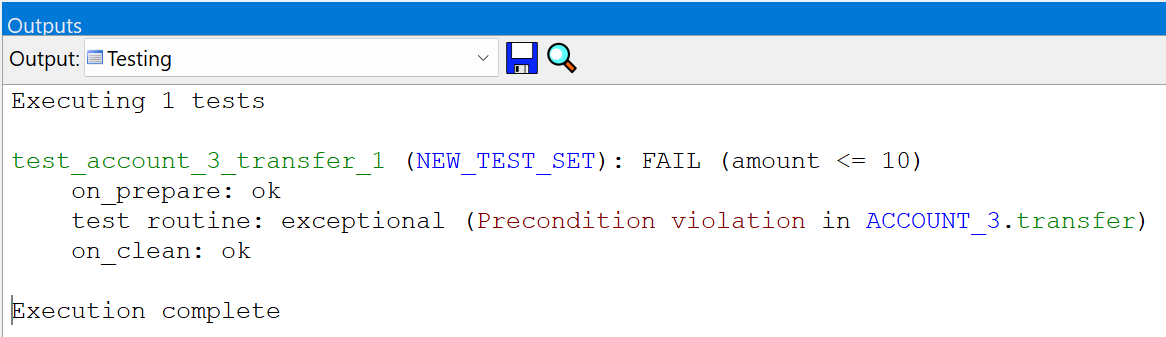}
  \label{fig: account_3_test_result}}
  \caption{(a) Verification result of \e{ACCOUNT_3} in AutoProof; (b) Testing result of \e{test_ACCOUNT_3_withdraw_1} in AutoTest}
  \label{fig:account_3}
\end{figure}

\newpage
\noindent \textbf{Variant 3 of \e{ACCOUNT}}
\label{account: variant 3}
\begin{itemize}[$\bullet$]
  \item {Fault injection}: after line 54, add a precondition \e{amount <= 10} for \e{deposit} to strengthen the precondition.
  \item {Resulting failure}: as shown in Figure  \ref{fig: account_3_proof_result}, the injected fault results in a failure of \e{transfer} --- it does not satisfy the new precondition \e{amount <= 10} when calling \e{deposit}.
  \item {Cause of the failure}: the inconsistency of specification between a supplier routine \e{deposit} and its client routine \e{transfer}: when the precondition of a routine is changed, its client routine should be changed accordingly. In this example, the upper limit of \e{transfer} should be consistent with the upper limit of \e{deposit}. In other words, the amount of money in a transfer operation should not exceed the maximum amount that is permitted in a deposit operation.
  \item Proof time: 0.248 sec
  \item Test generation time: 0.212 sec
  \item Resulting test case: Figure  \ref{fig: account_3_test_case} shows the test case from Proof2Test, which calls \e{transfer} with input --- \e{Current.balance = -2147483599}, \e{Current.credit_limit} \e{= -2147483632}, \e{amount = 33}, \e{other.balance = 7719}, \e{other.credit_limit = -2147481211}.
  \item Testings result: running the test case in raises the failure of precondition violation \e{amount <= 10} of \e{deposit} (same failure in the proof), as shown in Figure  \ref{fig: account_3_test_result}.
  \item Comment: during the execution of the test, when the program calls \e{other.deposit} from \e{transfer}, the input for \e{deposit} is \e{amount} = 33, which violates the precondition of \e{deposit} and demonstrates the problem.
\end{itemize}

\begin{figure}
\centering
\begin{lstlisting}[captionpos=b, basicstyle=\fontsize{0.27cm}{0.27cm}]
	test_ACCOUNT_3_transfer_1
		local
			current_object: ACCOUNT_3
			amount: INTEGER_32
			other: ACCOUNT_3
		do
			create current_object.make
			create other.make
			{P_INTERNAL}.set_integer_field_ (``balance'', current_object, (-2147483599))
				-- current_object.balance = (-2147483599)
			{P_INTERNAL}.set_integer_field_ (``credit_limit'', current_object, (-2147483632))
				-- current_object.credit_limit = (-2147483632)
			amount := 33
			{P_INTERNAL}.set_integer_field_ (``balance'', other, 7719)
				-- other.balance = 7719
			{P_INTERNAL}.set_integer_field_ (``credit_limit'', other, (-2147481211))
				-- other.credit_limit = (-2147481211)
			current_object.transfer (amount, other)
		end
\end{lstlisting}
\caption{Test case generated by Proof2Test}
\label{fig: account_3_test_case}
\end{figure}

\begin{figure}
  \centering
  \subfigure[]{
  \includegraphics[width=5.2in]{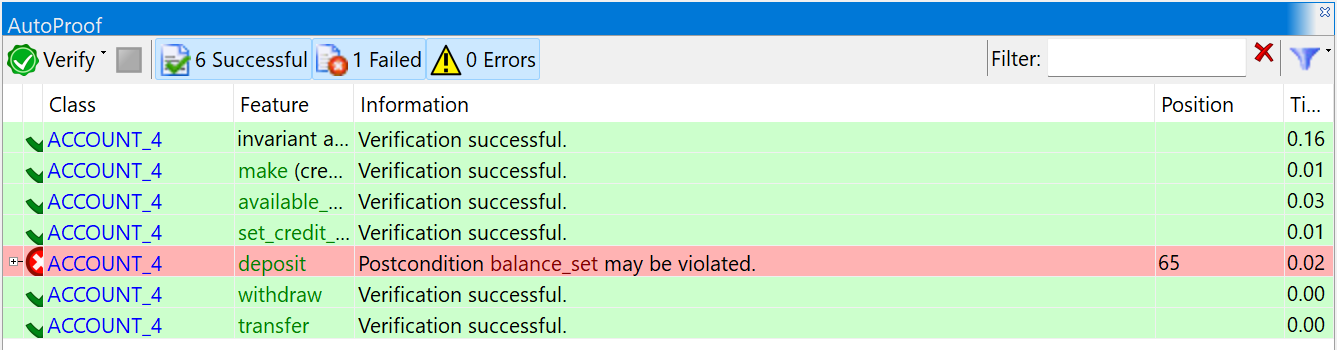}
  \label{fig: account_4_proof_result}
  }
  \subfigure[]{
  \includegraphics[width=4.2in]{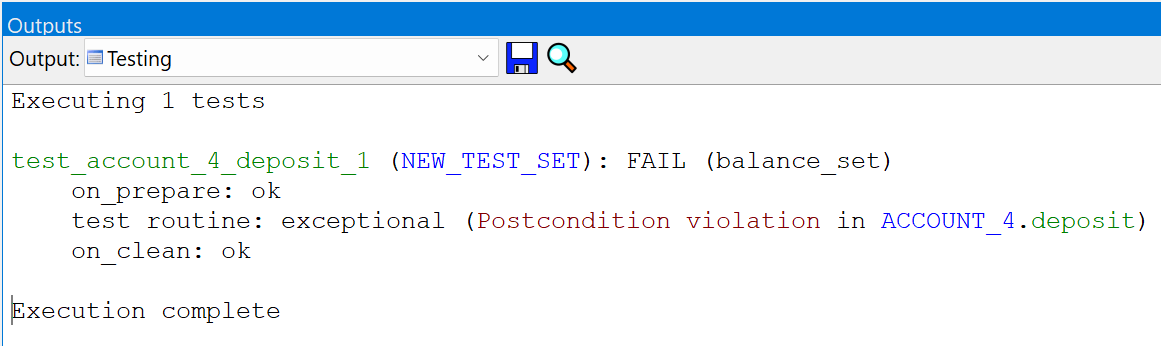}
  \label{fig: account_4_test_result}}
  \caption{(a) Verification result of \e{ACCOUNT_4} in AutoProof; (b) Testing result of \e{test_ACCOUNT_4_withdraw_1} in AutoTest}
  \label{fig:account_4}
\end{figure}

\newpage
\noindent \textbf{Variant 4 of \e{ACCOUNT}}
\label{account: variant 4}
\begin{itemize}[$\bullet$]
  \item {Fault injection}: at line 56, change the body of \e{deposit} from ``\e{balance := balance} \e{+ amount}'' into ``\e{balance := balance - amount}''.
  \item {Resulting failure}: as shown in Figure  \ref{fig: account_4_proof_result}, the postcondition \e{balance_set} is violated.
  \item {Cause of the failure}: this failure is similar to the failure in Variant 1, which results from the inconsistency between the implementation of \e{deposit} and its postcondition.
  \item Proof time: 0.241 sec
  \item Test generation time: 0.202 sec
  \item Resulting test case: Figure  \ref{fig: account_4_test_case} shows the test case from Proof2Test, which calls \e{deposit} with input --- \e{balance = 28101}, \e{credit_limit = 0}, \e{amount = 1}.
  \item Testings result: as shown in Figure  \ref{fig: account_4_test_result}, running the test raises an exception of postcondition violation of \e{balance_set}, which corresponds to the same failure in the proof.
  \item Comment: this Variant of \e{ACCOUNT} is similar to Variant 1; the values in the test input does not contain any specific meaning; running \e{deposit} with any valid test input would lead to the same contract violation.
\end{itemize}

\begin{figure}
\centering
\begin{lstlisting}[captionpos=b, basicstyle=\fontsize{0.27cm}{0.27cm}]
	test_ACCOUNT_4_deposit_1
		local
			current_object: ACCOUNT_4
			amount: INTEGER_32
		do
			create current_object.make
			{P_INTERNAL}.set_integer_field_ (``balance'', current_object, 28101)
				-- current_object.balance = 28101
			{P_INTERNAL}.set_integer_field_ (``credit_limit'', current_object, 0)
				-- current_object.credit_limit = 0
			amount := 1
			current_object.deposit (amount)
		end
\end{lstlisting}
\caption{Test case from failed proof of \e{balance_set}}
\label{fig: account_4_test_case}
\end{figure}

\begin{figure}
  \centering
  \subfigure[]{
  \includegraphics[width=5.4in]{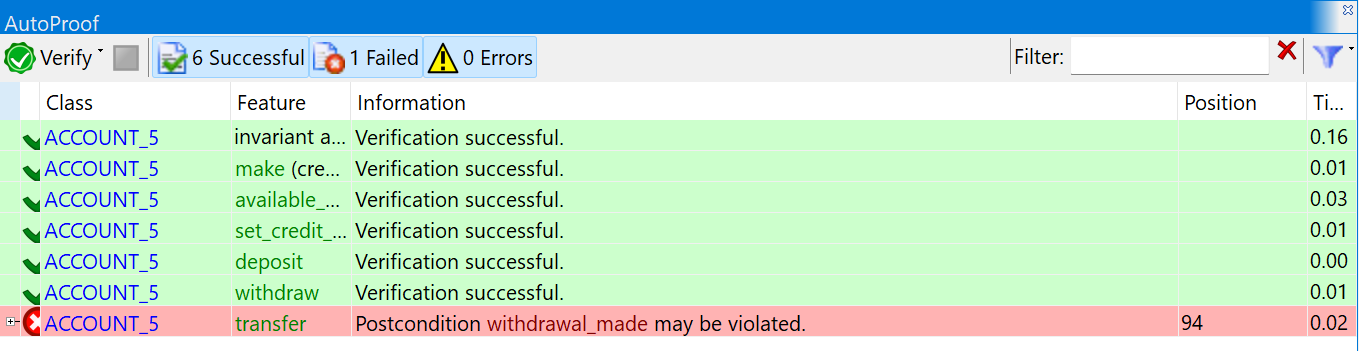}
  \label{fig: account_5_proof_result}
  }
  \subfigure[]{
  \includegraphics[width=4.2in]{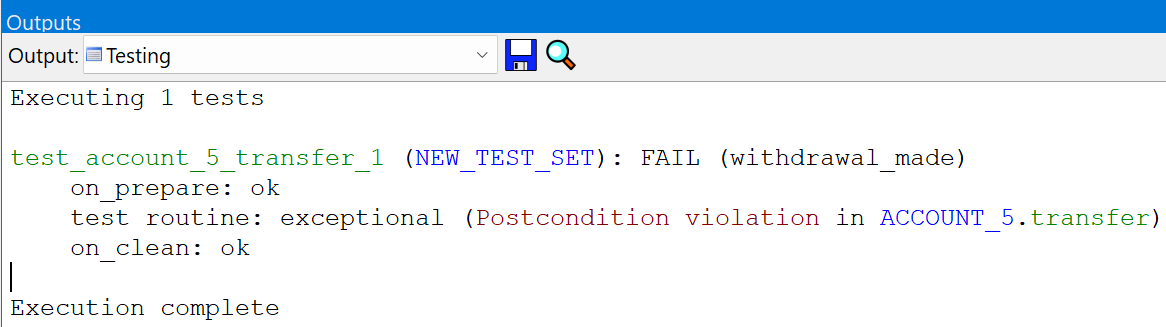}
  \label{fig: account_5_test_result}}
  \caption{(a) Verification result of \e{ACCOUNT_5} in AutoProof; (b) Testing result of \e{test_ACCOUNT_5_withdraw_1} in AutoTest}
  \label{fig:account_5}
\end{figure}

\newpage
\noindent \textbf{Variant 5 of \e{ACCOUNT}}
\label{account: variant 5}
\begin{itemize}[$\bullet$]
  \item {Fault injection}:  at line 87, remove the precondition \e{other} $\neq$ \e{Current} of \e{transfer}.
  \item {Resulting failure}: as shown in Figure  \ref{fig: account_5_proof_result}, the fault injection leads to the violation of postcondition \e{withdrawal_made} when verifying \e{transfer}.
  \item {Cause of the failure}: the precondition of \e{transfer} is too weak; it should exclude the case where an account transfers money to itself.
  \item Proof time: 0.243 sec
  \item Test generation time: 0.208 sec
  \item Resulting test case: Figure  \ref{fig: account_5_test_case} shows the test case, which calls \e{transfer} with input --- \e{balance = -2147481210}, \e{credit_limit = -2147482752}, \e{amount = 1542}, and \e{other} is an alias of \e{Current} (line 13).
  \item Testings result: as presented in Figure  \ref{fig: account_5_test_result}, running the test case raises the failure of violation of postcondition \e{withdrawal_made} of \e{transfer}, which is the same as the proof failure in AutoProof.
\end{itemize}

%

\begin{figure}
\centering
\begin{lstlisting}[captionpos=b, basicstyle=\fontsize{0.27cm}{0.27cm}]
	test_ACCOUNT_5_transfer_1
		local
			current_object: ACCOUNT_5
			amount: INTEGER_32
			other: ACCOUNT_5
		do
			create current_object.make
			{P_INTERNAL}.set_integer_field_ (``balance'', current_object, (-2147481210))
				-- current_object.balance = (-2147481210)
			{P_INTERNAL}.set_integer_field_ (``credit_limit'', current_object, (-2147482752))
				-- current_object.credit_limit = (-2147482752)
			amount := 1542
			other := current_object
			current_object.transfer (amount, other)
		end
\end{lstlisting}
\caption{Test case from failed proof of \e{withdrawal_made}}
\label{fig: account_5_test_case}
\end{figure}

\begin{figure}
  \centering
  \subfigure[]{
  \includegraphics[width=5.4in]{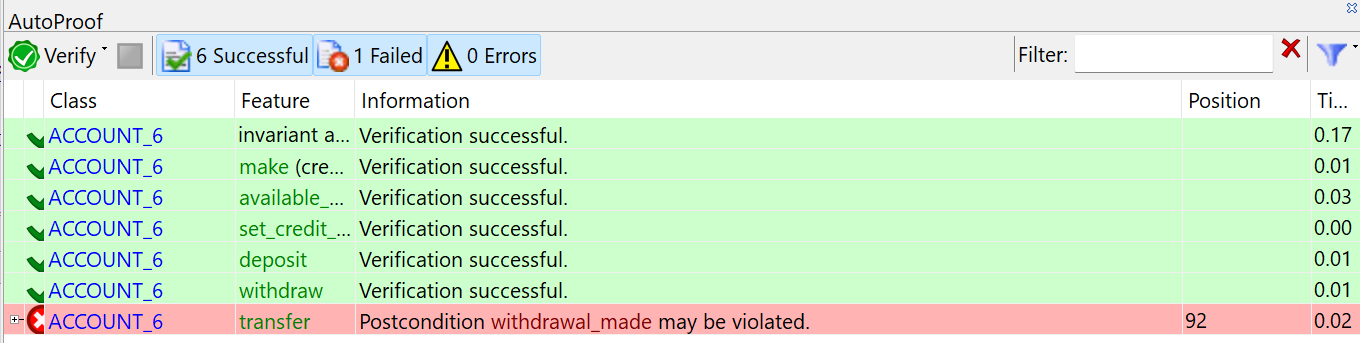}
  \label{fig: account_6_proof_result}
  }
  \subfigure[]{
  \includegraphics[width=3in]{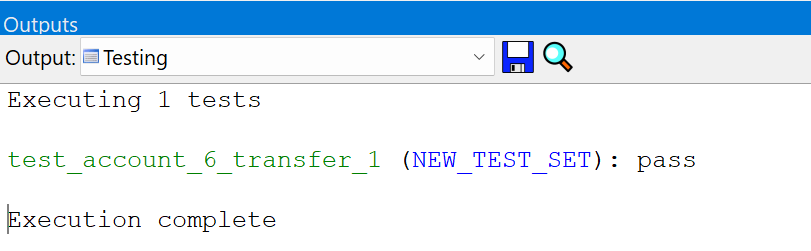}
  \label{fig: account_6_test_result}}
  \caption{(a) Verification result of \e{ACCOUNT_6} in AutoProof; (b) Testing result of \e{test_ACCOUNT_6_withdraw_1} in AutoTest}
  \label{fig:account_6}
\end{figure}

\newpage
\noindent \textbf{Variant 6 of \e{ACCOUNT}}
\label{account: variant 6}
\begin{itemize}[$\bullet$]
  \item {Fault injection}: at line 73, remove the postcondition \e{balance_set} of \e{withdraw}.
  \item {Resulting failure}: as shown in Figure  \ref{fig: account_6_proof_result}, the injected fault results in the violation of postcondition \e{withdrawal_made} when verifying \e{transfer}.
  \item {Cause of the failure}: the postcondition of \e{withdraw} is incomplete --- not strong enough to represent the functionality of \e{withdraw}; as a result, when reasoning about the correctness of its client routine \e{transfer}, the prover is not able to establish the postcondition \e{withdrawal_made} of \e{transfer}, which is related to the functionality of \e{withdraw}.
  \item Proof time: 0.253 sec
  \item Test generation time: 0.214 sec
  \item Resulting test case: Figure  \ref{fig: account_6_test_case} shows the test case, which calls \e{transfer} with input --- \e{Current.balance = -2147475928}, \e{Current.credit_limit = -2147475929}, \e{amount = 0}, \e{other.balance = 0}, and \e{other.credit_limit = 0}
  \item Testings result: as shown in Figure  \ref{fig: account_6_test_result}, the execution of the test terminates with no contract violation; this is because when verifying a client routine, AutoProof uses the postconditions of the involved supplier routines, instead of their bodies, to represent their functional behaviors; in this example, as the postcondition of \e{withdraw} does not strong enough to express its functionality (\e{balance} should be deduced by \e{amount}), AutoProof fails to establish the corresponding postcondition \e{withdrawal_made} of \e{transfer}; in other words, the counterexample, from which the test input is extracted, is not a real ``counterexample''; but the successfulness of the testing result reveals the weakness of specifications in the relevant routines.
\end{itemize}

\begin{figure}
\centering
\begin{lstlisting}[captionpos=b, basicstyle=\fontsize{0.27cm}{0.27cm}]
	test_ACCOUNT_6_transfer_1
		local
			current_object: ACCOUNT_6
			amount: INTEGER_32
			other: ACCOUNT_6
		do
			create current_object.make
			create other.make
			{P_INTERNAL}.set_integer_field_ (``balance'', current_object, (-2147475928))
				-- current_object.balance = (-2147475928)
			{P_INTERNAL}.set_integer_field_ (``credit_limit'', current_object, (-2147475929))
				-- current_object.credit_limit = (-2147475929)
			amount := 0
			{P_INTERNAL}.set_integer_field_ (``balance'', other, 0)
				-- other.balance = 0
			{P_INTERNAL}.set_integer_field_ (``credit_limit'', other, 0)
				-- other.credit_limit = 0
			current_object.transfer (amount, other)
		end
\end{lstlisting}
\caption{Test case generated by Proof2Test}
\label{fig: account_6_test_case}
\end{figure}

\begin{figure}
  \centering
  \subfigure[]{
  \includegraphics[width=5.4in]{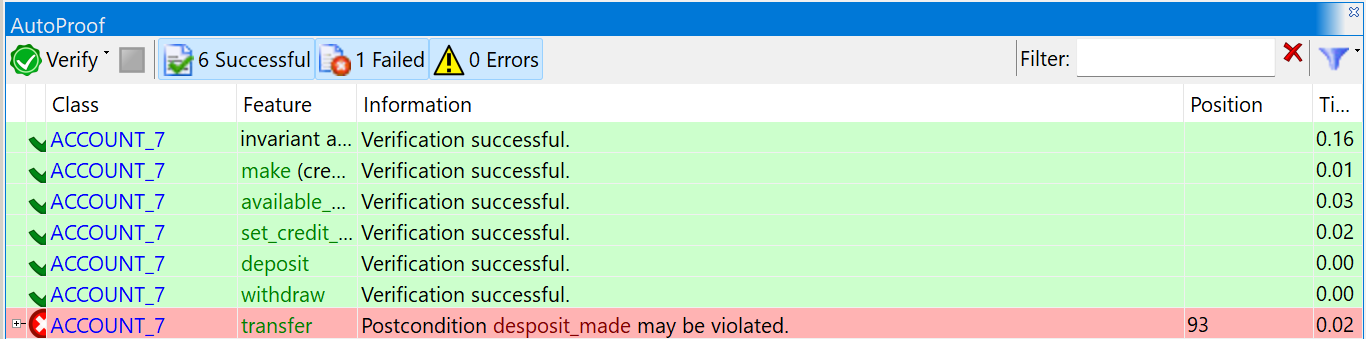}
  \label{fig: account_7_proof_result}
  }
  \subfigure[]{
  \includegraphics[width=3.2in]{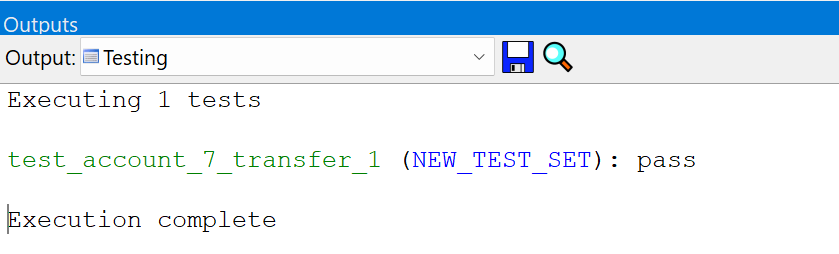}
  \label{fig: account_7_test_result}}
  \caption{(a) Verification result of \e{ACCOUNT_7} in AutoProof; (b) Testing result of \e{test_ACCOUNT_7_withdraw_1} in AutoTest}
  \label{fig:account_7}
\end{figure}

\newpage
\noindent \textbf{Variant 7 of \e{ACCOUNT}}
\label{account: variant 7}
\begin{itemize}[$\bullet$]
  \item {Fault injection}: at line 60, remove the postcondition \e{balance_set} of \e{deposit}.
  \item {Resulting failure}: the injected fault, as shown in Figure  \ref{fig: account_7_proof_result}, results in the violation of postcondition \e{deposit_made} when verifying \e{transfer}.
  \item {Cause of the failure}: similar to the previous failure (in Variant 6), this failure of \e{transfer} is due to the weakness of the postcondition of its supplier class \e{deposit} -- the postcondition is not strong enough to represent the functionality of \e{deposit}.
  \item Proof time: 0.244 sec
  \item Test generation time: 0.223 sec
  \item Resulting test case: Figure  \ref{fig: account_7_test_case} shows the test from Proof2Test, which calls \e{transfer} with input --- \e{Current.balance = 0}, \e{Current.credit_limit = 0}, \e{amount = 0}, \e{other.balance = 0}, and \e{other.credit_limit = -7720}.
  \item Testings result: as shown in Figure  \ref{fig: account_7_test_result}, execution of the test case terminates with no exception raised; similar to the failure in Variant 6, as the postcondition of the supplier routine \e{deposit} is too weak to describe its functional behavior (\e{balance} should be increased by \e{amount}), AutoProof fails to establish the postcondition \e{deposit_made} of \e{transfer}, requiring that \e{balance} of the \e{other} object should be increased by \e{amount}.
\end{itemize}

\begin{figure}
\centering
\begin{lstlisting}[captionpos=b, basicstyle=\fontsize{0.27cm}{0.27cm}]
	test_ACCOUNT_7_transfer_1
		local
			current_object: ACCOUNT_7
			amount: INTEGER_32
			other: ACCOUNT_7
		do
			create current_object.make
			create other.make
			{P_INTERNAL}.set_integer_field_ (``balance'', current_object, 0)
				-- current_object.balance = 0
			{P_INTERNAL}.set_integer_field_ (``credit_limit'', current_object, 0)
				-- current_object.credit_limit = 0
			amount := 0
			{P_INTERNAL}.set_integer_field_ (``balance'', other, 0)
				-- other.balance = 0
			{P_INTERNAL}.set_integer_field_ (``credit_limit'', other, (-7720))
				-- other.credit_limit = (-7720)
			current_object.transfer (amount, other)
		end
\end{lstlisting}
\caption{Test case from the failed proof of \e{deposit_made}}
\label{fig: account_7_test_case}
\end{figure}

\newpage
\subsection*{A.2 Results for class \e{LINEAR_SEARCH}}
\label{results for linear search}

\begin{figure}[htbp]
\centerline{{\includegraphics[width=4in]{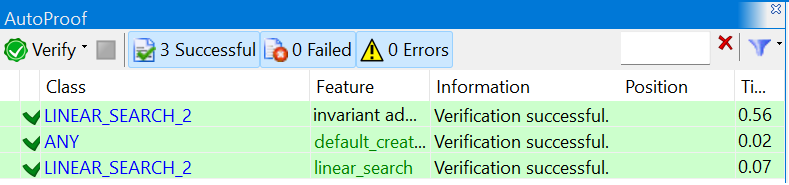}
}}
\caption{Proof result of \e{LINEAR_SEARCH} in AutoProof}
\label{fig: linear_search_proof_result}
\end{figure}

The \e{LINEAR_SEARCH} class, which is displayed below, implements a function that returns the index of a given integer `\e{value}' in an integer array `\e{a}' using linear search starting from beginning of the array; if the `\e{value}' is not found in `\e{a}', the function returns the value ``\e{a.count + 1}'' (\e{a.count} represents the number of elements in \e{a}). Figure \ref{fig: linear_search_proof_result} shows the verification result of \e{LINEAR_SEARCH}, which indicates the complete correctness of its functionality. 4 variants of \e{LINEAR_SEARCH} are produced based on the correct version and are discussed below.

\begin{lstlisting}[captionpos=b,  basicstyle=\fontsize{0.27cm}{0.27cm}]
class
    LINEAR_SEARCH

feature -- Basic operations
    linear_search (a: SIMPLE_ARRAY [INTEGER]; value: INTEGER): INTEGER
        require
            array_not_empty: a.count > 0
        do
            from
                Result := 1
            invariant
                result_in_bound: 1 <= Result and Result <= a.count + 1
                not_present_so_far: across 1 |..| (Result - 1) as i all a.sequence [i] /= value end
            until
                Result = $ $  a.count + 1 $ $  or else a [Result] = value
            loop
                Result := Result + 1
            variant
                a.count - Result + 1
            end
        ensure
            result_in_bound: 1 <= Result and Result <= a.count + 1
            present: a.sequence.has (value) = (Result <= a.count)
            found_if_present: (Result <= a.count) implies a.sequence [Result] = value
            first_from_front: across 1 |..| (Result - 1) as i all a.sequence [i] /= value end
        end
end
\end{lstlisting}

\begin{figure}
  \centering
  \subfigure[]{
  \includegraphics[width=5.8in]{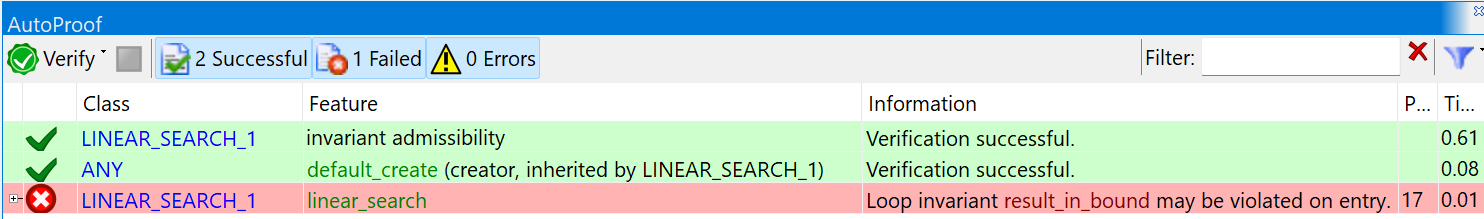}
  \label{fig: linear_search_1_proof_result}
  }
  \subfigure[]{
  \includegraphics[width=4.8in]{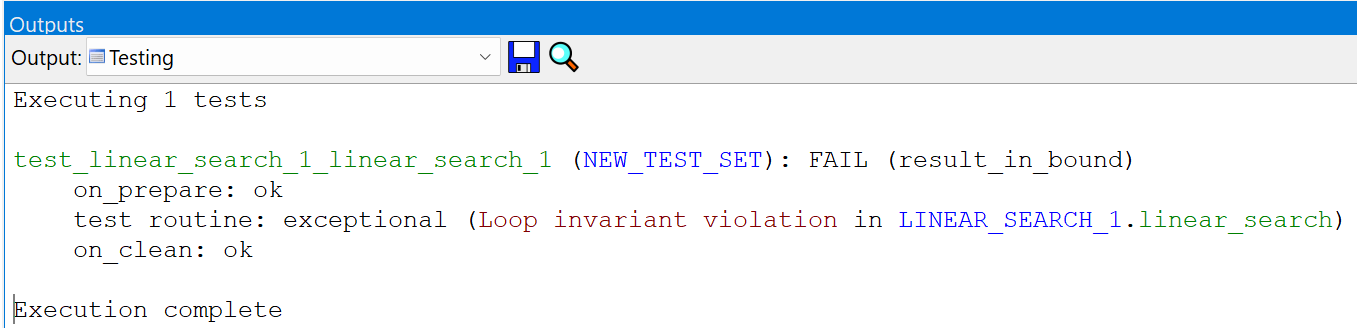}
  \label{fig: linear_search_1_test_result}}
  \caption{(a) Verification result of \e{LINEAR_SEARCH_1} in AutoProof; (b) Testing result of \e{test_LINEAR_SEARCH_1_linear_search_1} in AutoTest}
  \label{fig:linear_search_1}
\end{figure}

\newpage
\noindent \textbf{Variant 1 of \e{LINEAR_SEARCH}}
\label{linear search: variant 1}
\begin{itemize}[$\bullet$]
  \item {Fault injection}: at line 10, change the loop initialization from ``\e{Result := 1}'' into ``\e{Result := 0}''.
  \item {Resulting failure}: as shown in Figure  \ref{fig: linear_search_1_proof_result}, the injected fault leads to the violation of the loop invariant \e{result_in_bound} at the entry of the loop (after loop initialization).
  \item {Cause of the failure}: incorrect implementation of loop initialization.
  \item Proof time: 0.709 sec
  \item Test generation time: 0.332 sec
  \item Resulting test case: Figure  \ref{fig: linear_search_1_test_case} shows the test case from Proof2Test, which calls \e{linear_search} with input extracted from the corresponding counterexample: \e{a[1] = 0}, \e{a[2] = 0}, \e{value = (-2147475929)}.
  \item Testings result: as shown in Figure  \ref{fig: linear_search_1_test_result}, execution of the test case raises an exception of violation of loop invariant \e{result_in_bound}, which corresponds to the same proof failure.
  \item Comment: the test is useful as its execution demonstrates a specific case where the program goes to a failure state, violating the same contract as in the proof failure; the values in the test input, however, is not that meaningful to this failure, as running the program with any valid input would cause the same contract violation.
\end{itemize}

\begin{figure}
\centering
\begin{lstlisting}[captionpos=b,  basicstyle=\fontsize{0.27cm}{0.27cm}]
	test_LINEAR_SEARCH_1_linear_search_1
		local
			current_object: LINEAR_SEARCH_1
			a: SIMPLE_ARRAY[INTEGER_32]
			value: INTEGER_32
			linear_search_result: INTEGER_32
		do
			create current_object
			create a.make_empty
			a.force(0, 1)
			a.force(0, 2)

			value := (-2147475929)
			linear_search_result := current_object.linear_search (a, value)
		end
\end{lstlisting}
\caption{Test case from failed proof of \e{result_in_bound}}
\label{fig: linear_search_1_test_case}
\end{figure}

\begin{figure}
  \centering
  \subfigure[]{
  \includegraphics[width=5.4in]{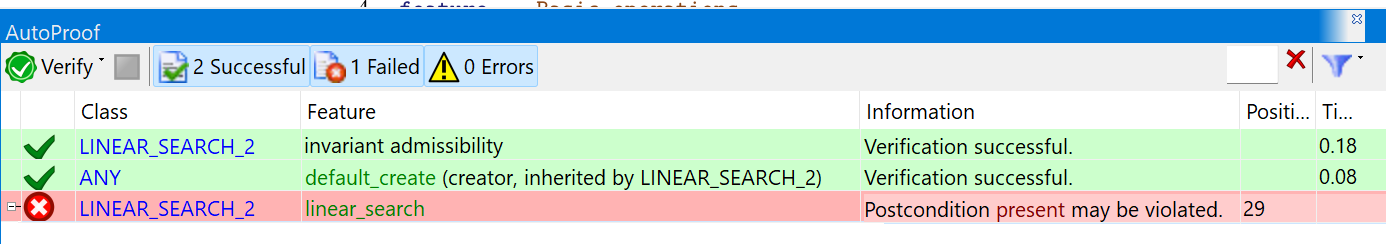}
  \label{fig: linear_search_2_proof_result}
  }
  \subfigure[]{
  \includegraphics[width=4.8in]{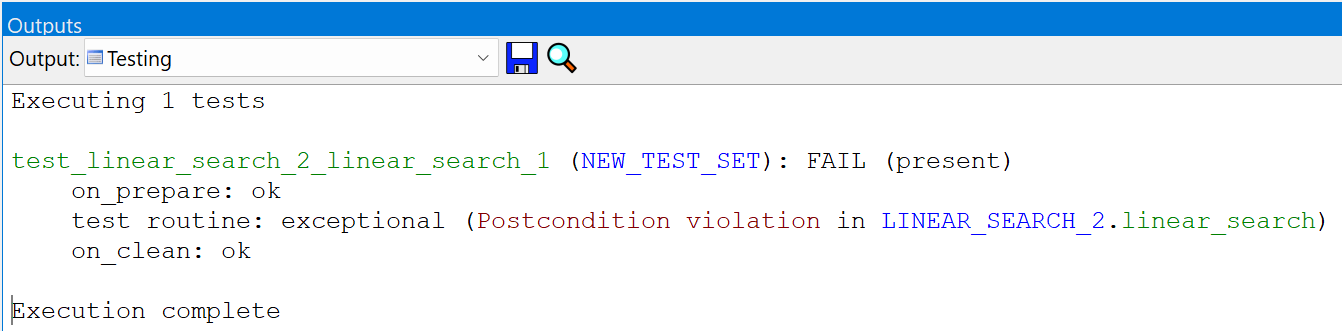}
  \label{fig: linear_search_2_test_result}}
  \caption{(a) Verification result of \e{LINEAR_SEARCH_2} in AutoProof; (b) Testing result of \e{test_LINEAR_SEARCH_2_linear_search_1} in AutoTest}
  \label{fig:linear_search_2}
\end{figure}

\newpage
\noindent \textbf{Variant 2 of \e{LINEAR_SEARCH}}
\label{linear search: variant 2}
\begin{itemize}[$\bullet$]
  \item {Fault injection}: at line 12, change the left part of the exit condition from ``\e{Result = a.count} \e{+ 1}'' into ``\e{Result = a.count}''.
  \item {Resulting failure}: as shown in Figure  \ref{fig: linear_search_2_proof_result}, the injected fault results in the violation of the postcondition \e{present}.
  \item {Cause of the failure}: incorrect exit condition (the condition for a loop to terminate).
  \item Proof time: 0.283 sec
  \item Test generation time: 0.373 sec
  \item Resulting test case: Figure  \ref{fig: linear_search_2_test_case} shows the test case, which calls \e{linear_search} with input extracted from the corresponding counterexample: \e{a[1] = 0, a[2] = 0}, \e{value = (-2147475282)}
  \item Testings result: as shown in Figure  \ref{fig: linear_search_2_test_result}, execution of the test case raises an exception of violation postcondition \e{present}, which corresponds to the same proof failure.
  \item Comment: during the execution of the test, the program terminates after 1 iteration with \e{Result = 2}; this leads to the violation of the postcondition \e{present} --- in the equality expression, the left-hand segment \e{a.sequence.has (value)} is false, as \e{value} does not match to any element of the input array \e{a}, while the right-hand segment \e{Result} $\leq$ \e{a.count} is true (\e{Result = 2} and \e{a.count =2}).
\end{itemize}

\begin{figure}
\centering
\begin{lstlisting}[captionpos=b,  basicstyle=\fontsize{0.27cm}{0.27cm}]
	test_LINEAR_SEARCH_2_linear_search_1
		local
			current_object: LINEAR_SEARCH_2
			a: SIMPLE_ARRAY [INTEGER_32]
			value: INTEGER_32
			linear_search_result: INTEGER_32
		do
			create current_object
			create a.make_empty
			a.force(0, 1)
			a.force(0, 2)

			value := (-2147475282)
			linear_search_result := current_object.linear_search (a, value)
		end
\end{lstlisting}
\caption{Test case from failed proof of \e{present}}
\label{fig: linear_search_2_test_case}
\end{figure}

\begin{figure}
  \centering
  \subfigure[]{
  \includegraphics[width=5.4in]{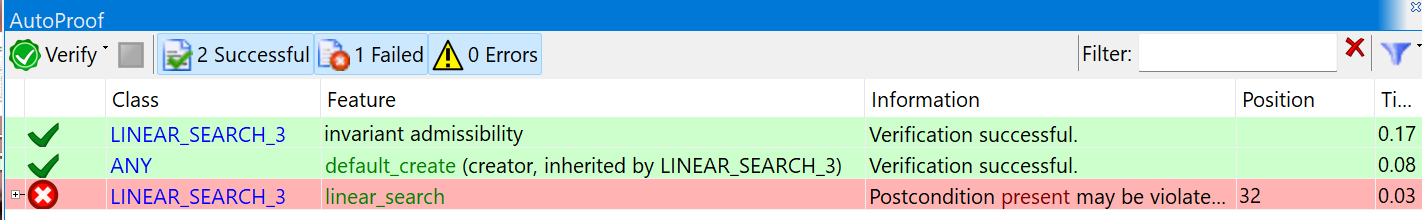}
  \label{fig: linear_search_3_proof_result}
  }
  \subfigure[]{
  \includegraphics[width=3.8in]{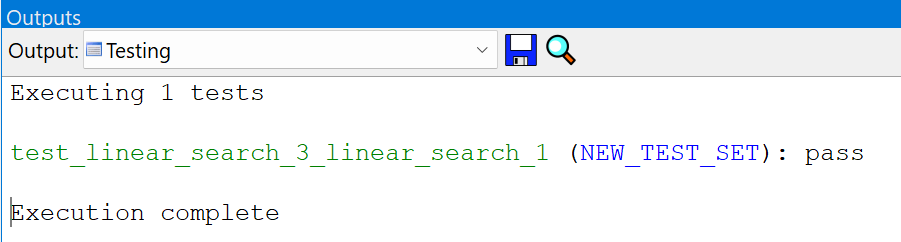}
  \label{fig: linear_search_3_test_result}}
  \caption{(a) Verification result of \e{LINEAR_SEARCH_3} in AutoProof; (b) Testing result of \e{test_LINEAR_SEARCH_3_binary_search_1} in AutoTest}
  \label{fig:linear_search_3}
\end{figure}

\newpage
\noindent \textbf{Variant 3 of \e{LINEAR_SEARCH}}
\label{linear search: variant 3}
\begin{itemize}[$\bullet$]
  \item {Fault injection}: at line 13, remove the loop invariant \e{not_present_so_far}.
  \item {Resulting failure}: as shown in Figure  \ref{fig: linear_search_3_proof_result}, the injected fault causes the violation of the postcondition \e{present}.
  \item {Cause of the failure}: weakness/incompleteness of loop invariant.
  \item Proof time: 0.280 sec
  \item Test generation time: 0.344 sec
  \item Resulting test case: Figure  \ref{fig: linear_search_3_test_case} shows the test case from Proof2Test, which calls \e{linear_search} with input: \e{a[1] = 0,} \e{a[2] = -2147462410}, \e{value = -2147462410}.
  \item Testings result: as shown in Figure  \ref{fig: linear_search_3_test_result}, execution of the test case does not raise any exception.
  \item Comment: when trying to verify postcondition \e{present}, the prover uses the loop invariant, instead of the loop body, to represent the behaviors of the loop; if the loop invariant is not strong enough to express the functionality of the loop, as in this example, the prover is not able to establish the relevant postcondition; in this case, the counterexample (from which the test is extracted) is not a real ``counterexample'', as it does not reveal the fault in the implementation; the passing test indicates that the proof failure is caused by the weakness of the auxiliary specification (the loop invariant), not the implementation.
\end{itemize}

%

\begin{figure}
\centering
\begin{lstlisting}[captionpos=b,  basicstyle=\fontsize{0.27cm}{0.27cm}]
	test_LINEAR_SEARCH_3_linear_search_1
		local
			current_object: LINEAR_SEARCH_3
			a: SIMPLE_ARRAY[INTEGER_32]
			value: INTEGER_32
			linear_search_result: INTEGER_32
		do
			create current_object
			create a.make_empty
			a.force(0, 1)
			a.force((-2147462410), 2)

			value := (-2147462410)
			linear_search_result := current_object.linear_search (a, value)
		end
\end{lstlisting}
\caption{Test case from failed proof of \e{present}}
\label{fig: linear_search_3_test_case}
\end{figure}

\begin{figure}
  \centering
  \subfigure[]{
  \includegraphics[width=5.2in]{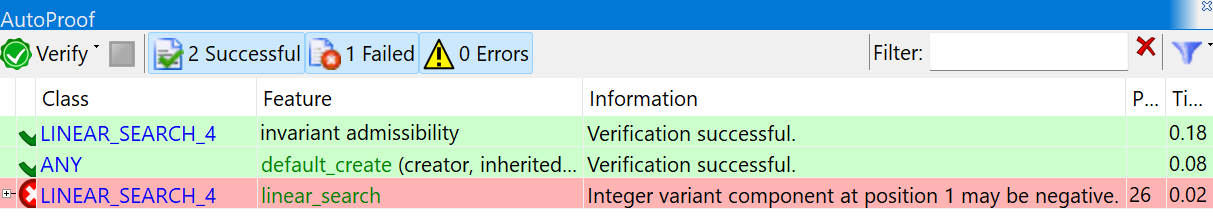}
  \label{fig: linear_search_4_proof_result}
  }
  \subfigure[]{
  \includegraphics[width=5.2in]{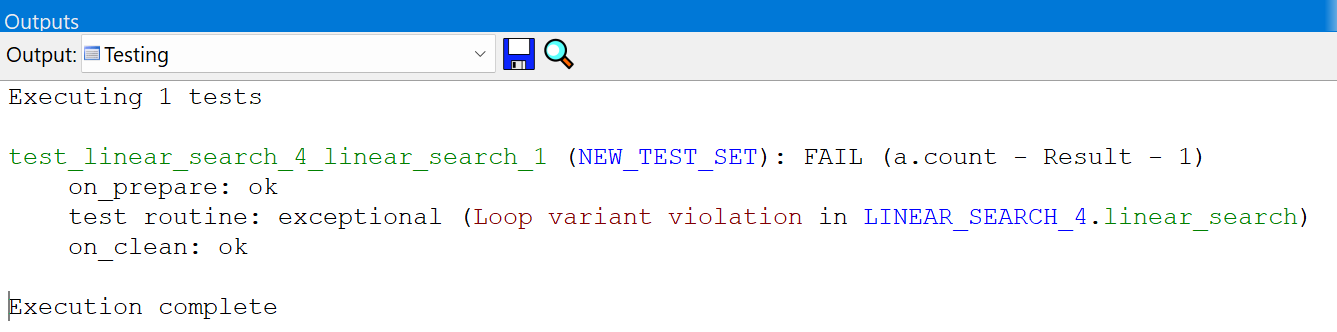}
  \label{fig: linear_search_4_test_result}}
  \caption{(a) Verification result of \e{LINEAR_SEARCH_4} in AutoProof; (b) Testing result of \e{test_LINEAR_SEARCH_4_binary_search_1} in AutoTest}
  \label{fig:linear_search_4}
\end{figure}

\newpage
\noindent \textbf{Variant 4 of \e{LINEAR_SEARCH}}
\label{linear search: variant 4}
\begin{itemize}[$\bullet$]
  \item {Fault injection}: change the loop variant at line 19 from ``\e{a.count - Result + 1}'' into ``\e{a.count - Result - 1}''.
  \item {Resulting failure}: as shown in Figure  \ref{fig: linear_search_4_proof_result}, the injected faults leads to the violation that ``the integer variant component at iteration 1 may be negative''.
  \item {Cause of the failure}: incorrect loop variant.
  \item Proof time: 0.279 sec
  \item Test generation time: 0.312 sec
  \item Resulting test case: Figure  \ref{fig: linear_search_4_test_case} shows the test case from Proof2Test, which calls \e{linear_search} with input extracted from the corresponding counterexample.
  \item Testings result: as shown in Figure  \ref{fig: linear_search_4_test_result}, execution of the test case raises an exception related to loop variant expression, which corresponds to the same failure in the verification.
  \item Comment: the test is useful as it is able to show how the value of variant varies at each iteration; the values in the test input, however, is not that meaningful, as any other valid test input will have the same effect.
  \end{itemize}

\begin{figure}
\centering
\begin{lstlisting}[captionpos=b,  basicstyle=\fontsize{0.27cm}{0.27cm}]
	test_LINEAR_SEARCH_4_linear_search_1
		local
			current_object: LINEAR_SEARCH_4
			a: SIMPLE_ARRAY[INTEGER_32]
			value: INTEGER_32
			linear_search_result: INTEGER_32
		do
			create current_object
			create a.make_empty
			a.force(0, 1)
			a.force(0, 2)
			a.force(0, 3)
			a.force(0, 4)
			a.force((-2147482506), 5)

			value := (-2147482505)
			linear_search_result := current_object.linear_search (a, value)
		end
\end{lstlisting}
\caption{Test case from failed proof of ``variant may be negative''}
\label{fig: linear_search_4_test_case}
\end{figure}

\end{document}